\documentclass[amsmath,amssymb,aps,prx,reprint,superscriptaddress]{revtex4-2}
\usepackage{amsmath}
\usepackage[table]{xcolor}
\usepackage{amsfonts,amssymb}
\usepackage{graphicx}
\usepackage{hyperref}
\usepackage{academicons}
\usepackage{multirow}
\usepackage[version=4]{mhchem}
\usepackage{makecell}
\usepackage{pgf}
\usepackage{booktabs}
\usepackage[english]{babel}

\newcommand{\CellMin}{0}
\newcommand{\CellMax}{0.4}
\colorlet{cmin}{green!10!white}  
\colorlet{cmed}{yellow!20!white} 
\colorlet{cmax}{red!20!white}    
\newcommand{\ColorCell}[1]{%
  \begingroup
  \pgfmathsetmacro{\val}{#1}%
  \pgfmathsetmacro{\norm}{(\val-\CellMin)/(\CellMax-\CellMin)}%
  \pgfmathsetmacro{\norm}{max(min(\norm,1),0)}%
  \pgfmathsetmacro{\pct}{\norm*100}%
  \pgfmathtruncatemacro{\pctint}{round(\pct)}%

  \ifnum\pctint<50
    \pgfmathtruncatemacro{\mix}{round(100 - 2*\pctint)}%
    \edef\temp{\noexpand\cellcolor{cmin!\mix!cmed}}%
  \else
    \pgfmathtruncatemacro{\mix}{round(200 - 2*\pctint)}%
    \edef\temp{\noexpand\cellcolor{cmed!\mix!cmax}}%
  \fi
  \temp #1%
  \endgroup
}

\begin{document}

\title{
Benchmarking foundation potentials against quantum chemistry methods for predicting molecular redox potentials}

\author{Yicheng Chen}
\affiliation{Department of Materials Science and Engineering, National University of Singapore, Singapore 117575, Singapore}

\author{Lixue Cheng}
\affiliation{Department of Chemistry, Hong Kong University of Science and Technology, Kowloon, Hong Kong 999077, China}

\author{Yan Jing}
\affiliation{Department of Materials Science and Engineering, National University of Singapore, Singapore 117575, Singapore}

\author{Peichen Zhong}
\email[]{zhongpc@nus.edu.sg}
\affiliation{Department of Materials Science and Engineering, National University of Singapore, Singapore 117575, Singapore}


\date{\today}

\begin{abstract}
Computational high-throughput virtual screening is essential for identifying redox-active molecules for sustainable applications such as electrochemical carbon capture. A primary challenge in this approach is the high computational cost associated with accurate quantum chemistry calculations. Machine learning foundation potentials (FPs) trained on extensive density functional theory (DFT) calculations offer a computationally efficient alternative. Here, we benchmark the MACE-OMol-0 and UMA FPs against a hierarchy of DFT functionals for predicting experimental molecular redox potentials for both electron transfer (ET) and proton-coupled electron transfer (PCET) reactions. We find that these FPs achieve exceptional accuracy for PCET processes, rivaling their target DFT method. However, the performance is diminished for ET reactions, particularly for multi-electron transfers involving reactive ions that are underrepresented in the OMol25 training data, revealing a key out-of-distribution limitation. To overcome this, we propose an optimal hybrid workflow that uses the FPs for efficient geometry optimization and thermochemical analysis, followed by a crucial single-point DFT energy refinement and an implicit solvation correction. This pragmatic approach provides a robust and scalable strategy for accelerating high-throughput virtual screening in sustainable chemistry.
\end{abstract}



\maketitle
\section{Introduction}

\begin{figure*}[htbp]
    \centering
    \includegraphics[width=\linewidth]{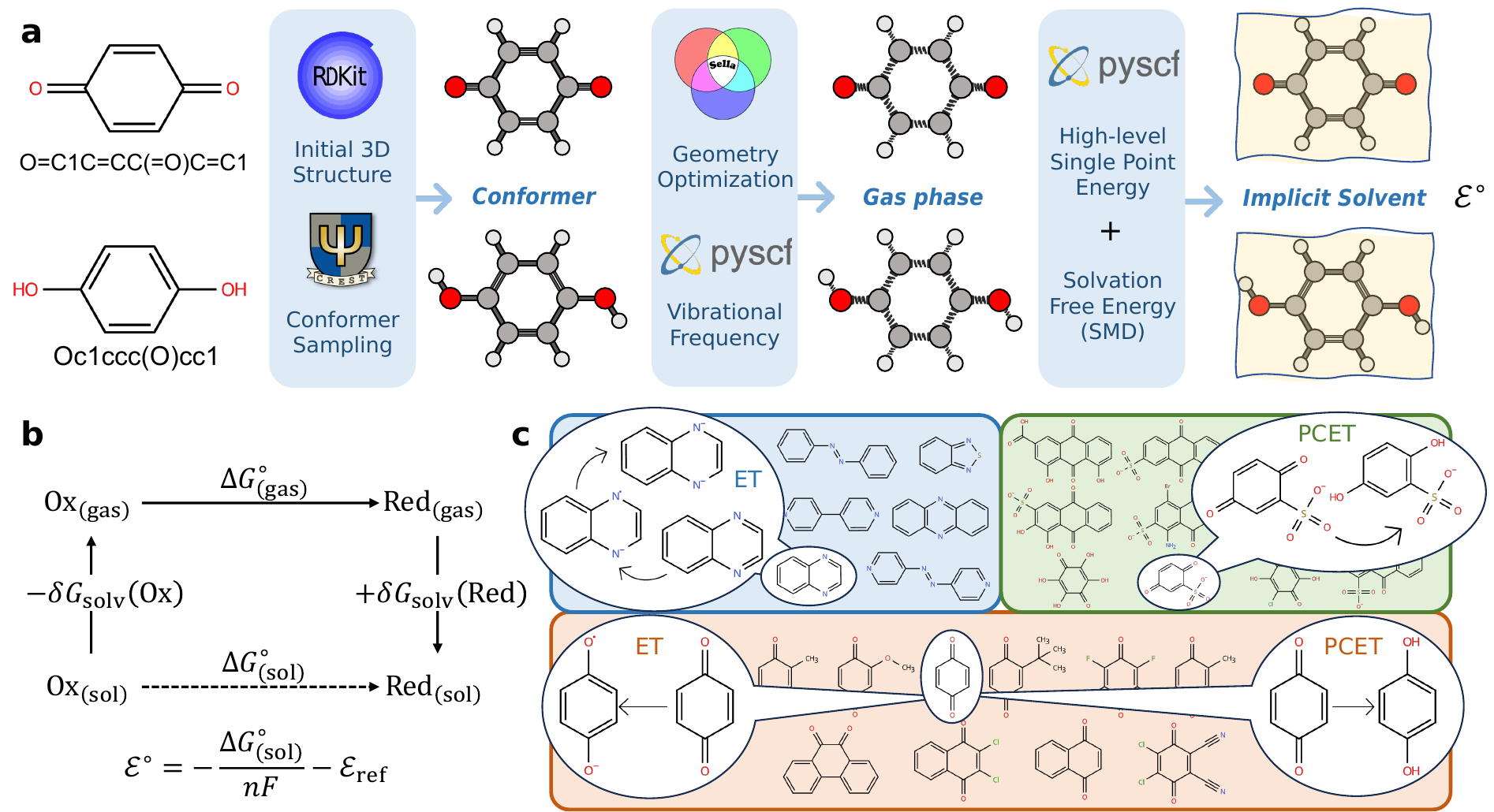}
    \caption{Overview of the computational workflow. (a) Key steps for calculating free energy, including conformer search, geometry optimization, vibrational frequency analysis, and single-point energy correction. (b) The Born-Haber cycle used to incorporate the solvation free energy ($\delta G_\text{solv}$) via the SMD implicit solvent model. (c) The three experimental datasets used for benchmarking, covering electron transfer (ET) and proton-coupled electron transfer (PCET) reactions.}
    \label{fig:workflow}
\end{figure*}

The development of efficient and scalable materials for \ce{CO2} capture is a cornerstone of advancing sustainable technologies \cite{chu_carbon_2009, lin_scalable_2021, siegelman_porous_2021,  zhou_carbon_2024}. Flow-based electrochemical systems offer a compelling alternative to traditional thermal and pressure-swing methods \cite{sharifian_electrochemical_2021, diederichsen_electrochemical_2022}, 
primarily due to their lower energy requirements and potential for integration with renewable energy sources \cite{wang_materials_2025}. The operational principle of these systems often relies on redox-active sorbent molecules \cite{voskian_faradaic_2019, diederichsen_solventfree_2022, li_lewis-base_2022} that facilitate \ce{CO2} capture and release through one of two primary mechanisms: direct binding of \ce{CO2} upon reduction of the sorbent with electron transfer (ET), or a proton-coupled electron transfer (PCET) reaction that generates hydroxide ions (\ce{OH-}) to capture \ce{CO2} in aqueous media \cite{jin_ph_2020, xie_lowenergy_2020, jing_electrochemically_2024}. The success of this approach is critically dependent on identifying sorbent candidates with precisely tuned redox potentials, ideally close to a relevant benchmark such as the oxygen reduction reaction, to ensure electrochemical reversibility in the presence of oxygen. 
Computational high-throughput screening has become an indispensable tool in materials discovery \cite{horton_accelerated_2025}. 
Quantum chemistry methods, such as Density Functional Theory (DFT) \cite{hohenberg_hk_1964, kohn_ks_1965}, enable the accurate and high-throughput prediction of redox potentials for many sorbent materials by calculating their free energies in reduced and oxidized states \cite{baik_computing_2002}
However, the high computational cost of DFT presents a significant bottleneck for screening the vast chemical space of potential sorbents. 

To address this challenge, the field has progressed along two parallel and complementary fronts. 
First, advancements in GPU-based computational infrastructure have significantly accelerated quantum chemistry methods (e.g., GPU4PySCF) \cite{wu_gpu4pyscf_2025}. 
This performance gain has made direct DFT calculations more tractable for large systems (e.g., enzyme catalysis \cite{li_accurate_2025}) and for generating large datasets used in training machine learning force fields for lithium ion battery liquid electrolytes \cite{gong_predictive_2025}. 
Concurrently, the advent of universal machine learning interatomic potentials (MLIPs), also termed foundation potentials (FPs), represents another paradigm shift \cite{chen_universal_2022, deng_chgnet_2023, batatia_mace_2023, yang_mattersim_2024, rhodes_orb-v3_2025, fu_learning_2025, bochkarev_graph_2024, zhang_graph_2025, kim_dataefficient_2025, wood_uma_2025}.  
These FPs, such as MACE-OMol-0 (hereinafter referred to as MACE-OMol) built with the high-order equivariant message passing neural network (MACE) \cite{batatia_mace_2023} and UMA (Universal Machine learning Atoms) \cite{wood_uma_2025}, are designed for broad applicability across chemical space.
MACE-OMol and UMA were trained on the OMol25 dataset \cite{levine_omol25_2025}. OMol25 is one of the largest and most diverse quantum chemistry resources available to date, comprising over 100 million DFT calculations performed at the $\omega$B97M-V/def2-TZVPD level of theory \cite{mardirossian_wb97mv_2016, rappoport_def2D_2010, weigend_def2_2005}. For example, MACE-OMol predicts energies with a mean absolute error (MAE) of 1.2 meV/atom and interatomic forces with an MAE of 10 meV/\AA\ relative to the reference DFT method. 
This remarkable accuracy suggests that such FPs could enable efficient molecular virtual screening campaigns without incurring the high computational cost of DFT.
Yet, their reliability for predicting derived electrochemical properties requires thorough validation \cite{vanzanten_benchmarking_2025}, which depends on subtle energy differences between distinct charge/spin and protonation states. 

In this work, we present a benchmark of two representative FPs -- MACE-OMol and UMA-s (with the OMol head) -- against a hierarchy of conventional quantum chemistry methods, including B3LYP \cite{becke_b3_1993,lee_lyp_1988,stephens_b3lyp_1994}, M06-2X \cite{zhao_m062x_2008}, $\omega$B97X \cite{chai_wb97x_2008} and $\omega$B97M \cite{mardirossian_wb97mv_2016}, along with DFT-D3 \cite{grimme_d3_2010,grimme_d3bj_2011} for dispersion correction. We evaluate its performance in calculating redox potentials for both direct ET and PCET reactions, providing insights into its readiness and the optimal approach to deploy FPs for predictive screening in sustainable chemistry applications.

\begin{table*}[tb]
    \centering
    \begin{tabular}{lcccccccc}
        \toprule
        OPT Method & B3LYP-D3(BJ) & M06-2X-D3 & $\omega$B97X-D3(BJ) & $\omega$B97M-D3(BJ) & \multicolumn{2}{c}{MACE-OMol} & \multicolumn{2}{c}{UMA-s} \\
        \midrule
        SP Method & B3LYP-D3(BJ) & M06-2X-D3 & $\omega$B97X-D3(BJ) & $\omega$B97M-D3(BJ) & MACE-OMol & $\omega$B97M-V & UMA-s & $\omega$B97M-V \\
        \midrule
        BPy (\ce{1e-}) & \ColorCell{0.144} & \ColorCell{0.016}& \ColorCell{0.022}& \ColorCell{0.088}& \ColorCell{0.169}& \ColorCell{0.045}& \ColorCell{0.089}& \ColorCell{0.089}\\
        QX (\ce{1e-}) & \ColorCell{0.119}& \ColorCell{0.049}& \ColorCell{0.060}& \ColorCell{0.023}& \ColorCell{0.072}& \ColorCell{0.001}& \ColorCell{0.032}& \ColorCell{0.003}\\
        BNSN (\ce{1e-}) & \ColorCell{0.152} & \ColorCell{0.095}& \ColorCell{0.129} & \ColorCell{0.066}& \ColorCell{0.190}& \ColorCell{0.002}& \ColorCell{0.072}& \ColorCell{0.021}\\
        AzB (\ce{1e-}) & \ColorCell{0.198}& \ColorCell{0.099}& \ColorCell{0.075}& \ColorCell{0.041}& \ColorCell{0.137}& \ColorCell{0.043}& \ColorCell{0.012}& \ColorCell{0.019}\\
        PhN (\ce{1e-}) & \ColorCell{0.112} & \ColorCell{0.088} & \ColorCell{0.073} & \ColorCell{0.045}& \ColorCell{0.113}& \ColorCell{0.007}& \ColorCell{0.031}& \ColorCell{0.045}\\
        AzPy (\ce{1e-}) & \ColorCell{0.150}& \ColorCell{0.016}& \ColorCell{0.034}& \ColorCell{0.014}& \ColorCell{0.287}& \ColorCell{0.077}& \ColorCell{0.065}& \ColorCell{0.048}\\
        \midrule
        BNSN (\ce{2e-}) & \ColorCell{0.470} & \ColorCell{0.441}& \ColorCell{0.369}& \ColorCell{0.442}& \ColorCell{4.015}& \ColorCell{0.237}& \ColorCell{3.587}& \ColorCell{0.195}\\
        AzB (\ce{2e-}) & \ColorCell{0.083}& \ColorCell{0.095}& \ColorCell{0.037}& \ColorCell{0.026}& \ColorCell{0.850}& \ColorCell{0.217}& \ColorCell{0.032}& \ColorCell{0.083}\\
        PhN (\ce{2e-}) & \ColorCell{0.164}& \ColorCell{0.173} & \ColorCell{0.057} & \ColorCell{0.108}& \ColorCell{2.279}& \ColorCell{0.647}& \ColorCell{2.142}& \ColorCell{0.584}\\
        AzPy (\ce{2e-}) & \ColorCell{0.127}& \ColorCell{0.171}& \ColorCell{0.078}& \ColorCell{0.113}& \ColorCell{0.663}& \ColorCell{0.015}& \ColorCell{0.036}& \ColorCell{0.042}\\
        \midrule
        MAE (\ce{1e-}) & \ColorCell{0.146}& \ColorCell{0.060}& \ColorCell{0.066}& \ColorCell{0.046}& \ColorCell{0.162}& \ColorCell{0.029}& \ColorCell{0.050}& \ColorCell{0.037}\\
        MAE (\ce{2e-}) & \ColorCell{0.211}& \ColorCell{0.220}& \ColorCell{0.135}& \ColorCell{0.172}& \ColorCell{1.952}& \ColorCell{0.279}& \ColorCell{1.449}& \ColorCell{0.226}\\
        Total MAE & \ColorCell{0.172}& \ColorCell{0.124}& \ColorCell{0.093}& \ColorCell{0.096}& \ColorCell{0.878}& \ColorCell{0.129}& \ColorCell{0.610}& \ColorCell{0.113} \\
        \bottomrule
    \end{tabular}
    \caption{
    Absolute errors in ET redox potentials for Test Set A (units in V). For DFT methods, geometries were optimized using the def2-SVPD basis set, followed by single-point energy calculations with the def2-TZVPD basis set.
    }
    \label{tab:lewis}
\end{table*}

\section{Methods}

\textbf{DFT calculations.}~
Our computational study was conducted using the open-source quantum chemistry package, PySCF 2.10.0 \cite{sun_pyscf_2020}. 
To manage the high computational cost, all DFT calculations were performed using the GPU4PySCF extension 1.4.3 \cite{li_gpu4pyscf_2025,wu_gpu4pyscf_2025}, which uses GPU-accelerated kernels to accelerate energy and gradient calculations for self-consistent field methods and implicit solvation. The DFT calculation settings included an SCF convergence threshold of $10^{-8}$ Hartree, a grid size of $(99, 590)$, density fitting for J/K integrals with auxiliary basis of def2-universal-JKFIT \cite{weigend_def2-jkfit_2008}, and all other parameters retained as default in GPU4PySCF.

\textbf{Solvation model.}~
To model the system within a continuous dielectric medium representative of flow-based electrochemical processes, we used the Solvation Model based on Density (SMD) as an implicit solvation model \cite{marenich_smd_2009}.
A key advantage of using the SMD solvation model is that its development is based on the optimized gas-phase structures. Given that the labels fit by MLIPs are typically gas-phase DFT results, SMD can serve as an external correction for solvation free energy.

This alignment enables the total solution-phase free energy $G_{(\text{sol})}$ to be decomposed into three additive components:
\begin{equation}\label{eq:free_energy}
    G_{(\text{sol})} = E_{(\text{g})} + \delta G_{(\text{g})} + \delta G_\text{solv}.
\end{equation}
Each component is computed at a distinct theoretical level: $\delta G_{(\text{g})}$ (gas-phase Gibbs free energy correction, accounting for thermostatistical effects) is obtained via geometry optimization and vibrational frequency analysis using low-cost methods; $E_{(\text{g})}$ (gas-phase single-point electronic energy) is calculated using higher-cost methods based on the optimized gas-phase structures; $\delta G_\text{solv}$ (solvation free energy) is calculated through M06-2X/6-31G(d) \cite{ditchfield_631gd_1971,hehre_631gd_1972,hariharan_631gd_1973,francl_631gd_1982,gordon_631gd_1982} which is compatible with the SMD model \cite{marenich_smd_2009,ribeiro_smd_eg_2011}. Specifically, we have
\begin{equation}\label{eq:solvation}
    \delta G_\text{solv} = E^\text{M06-2X}_{\text{(SMD)}} - E^\text{M06-2X}_{(\text{g})},
\end{equation}
where $E^\text{M06-2X}_{\text{(SMD)}}$ and $E^\text{M06-2X}_{(\text{g})}$ correspond to single-point energies under gas-phase and SMD solvated conditions, respectively.
This hybrid strategy offers a computationally tractable yet accurate method for modeling solvated molecular sorbents, utilizing the FP for geometrical optimization and frequency calculations.

\textbf{Redox potential.}~
The redox potential calculation relies on the free energy of oxidized and reduced molecules in the solvent. 
We constructed a Born-Haber cycle (Figure~\ref{fig:workflow}b) to derive the free energy difference between these two molecular states \cite{morris_born-fajans-haber_1969}.
For the general redox reaction \ce{Ox + n e^-} $\rightarrow$ Red in solvent, the standard reaction Gibbs free energy $\Delta G^\circ_{(\text{sol})}$ is
\begin{equation}
    \Delta G^\circ_{(\text{sol})} = G_{(\text{sol})}(\text{Red}) - G_{(\text{sol})}(\text{Ox}),
\end{equation}
where $G_{(\text{sol})}(\cdot)$ represents the Gibbs free energy in the solution calculated through Equation~\eqref{eq:free_energy}. The redox potential is given by
\begin{equation}\label{eq:potential}
    \mathcal{E}^\circ = -\frac{\Delta G^\circ_{(\text{sol})}}{nF} - \mathcal{E}_\text{ref},
\end{equation}
where $n$ is the number of electrons transferred in the reaction, $F$ is the Faraday constant, and $\mathcal{E}_\text{ref}$ represents the potential of the reference electrode.

The workflow for redox potentials is shown in Figure \ref{fig:workflow}a. First, to obtain molecular conformations, we used RDKit \cite{landrum_rdkit_2025} to generate the initial 3D structures, and then employed CREST \cite{pracht_crest_2024} to identify the most stable conformations with GFN2-xTB under GB/SA solvation model \cite{qiu_gbas_1997, bannwarth_gfn2xtb_2019}. 
Using these structures as starting points, we performed geometry optimizations and vibrational frequency analysis using several dispersion-corrected DFT functionals with def2-SVPD \cite{weigend_def2_2005,rappoport_def2D_2010} basis set, including B3LYP-D3(BJ) \cite{becke_b3_1993,lee_lyp_1988,stephens_b3lyp_1994,grimme_d3_2010,grimme_d3bj_2011}, M06-2X-D3 \cite{zhao_m062x_2008,grimme_d3_2010}, $\omega$B97X-D3(BJ) \cite{chai_wb97x_2008, najibi_wb97x/m-d3bj_2018} and $\omega$B97M-D3(BJ) \cite{mardirossian_wb97mv_2016,najibi_wb97x/m-d3bj_2018}, along with MACE-OMol FP. The geometry optimizations were performed using the Sella package 2.3.5 \cite{hermes_sella_2021}.

Given the optimized gas-phase structures, we calculated single-point electronic energy using the def2-TZVPD basis set \cite{weigend_def2_2005,rappoport_def2D_2010}, building on the same DFT functionals.
For MACE-OMol and UMA-s, we calculated their single-point energies using their target level of theory, i.e., $\omega$B97M-V/def2-TZVPD.
The final step involves calculating solvation free energies. Single-point energies under gas-phase and SMD solvated conditions were computed using M06-2X/6-31G(d), and solvation free energies were derived from Equation~\eqref{eq:solvation}.

\begin{table*}[tbp]
    \centering
    \begin{tabular}{lcccccccc}
        \toprule
        OPT Method & B3LYP-D3(BJ) & M06-2X-D3 & $\omega$B97X-D3(BJ) & $\omega$B97M-D3(BJ) & \multicolumn{2}{c}{MACE-OMol} & \multicolumn{2}{c}{UMA-s} \\
        \midrule
        SP Method & B3LYP-D3(BJ) & M06-2X-D3 & $\omega$B97X-D3(BJ) & $\omega$B97M-D3(BJ) & MACE-OMol & $\omega$B97M-V & UMA-s & $\omega$B97M-V \\
        \midrule
        AQDH12    & \ColorCell{0.092} & \ColorCell{0.076} & \ColorCell{0.121} & \ColorCell{0.014} & \ColorCell{0.097} & \ColorCell{0.088} & \ColorCell{0.098} & \ColorCell{0.096} \\
        AQDH14    & \ColorCell{0.030} & \ColorCell{0.045} & \ColorCell{0.082} & \ColorCell{0.047} & \ColorCell{0.154} & \ColorCell{0.151} & \ColorCell{0.150} & \ColorCell{0.148} \\
        AQDH15    & \ColorCell{0.182} & \ColorCell{0.156} & \ColorCell{0.197} & \ColorCell{0.066} & \ColorCell{0.048} & \ColorCell{0.040} & \ColorCell{0.045} & \ColorCell{0.040} \\
        AQDH18    & \ColorCell{0.044} & \ColorCell{0.047} & \ColorCell{0.013} & \ColorCell{0.144} & \ColorCell{0.242} & \ColorCell{0.240} & \ColorCell{0.215} & \ColorCell{0.213} \\
        AQDH26    & \ColorCell{0.011} & \ColorCell{0.046} & \ColorCell{0.002} & \ColorCell{0.134} & \ColorCell{0.232} & \ColorCell{0.226} & \ColorCell{0.210} & \ColorCell{0.208} \\
        AQDS27    & \ColorCell{0.127} & \ColorCell{0.102} & \ColorCell{0.143} & \ColorCell{0.005} & \ColorCell{0.129} & \ColorCell{0.136} & \ColorCell{0.144} & \ColorCell{0.147} \\
        AQDS15    & \ColorCell{0.386} & \ColorCell{0.350} & \ColorCell{0.364} & \ColorCell{0.227} & \ColorCell{0.157} & \ColorCell{0.150} & \ColorCell{0.151} & \ColorCell{0.152} \\
        AQDS18    & \ColorCell{0.251} & \ColorCell{0.245} & \ColorCell{0.207} & \ColorCell{0.078} & \ColorCell{0.027} & \ColorCell{0.020} & \ColorCell{0.030} & \ColorCell{0.027} \\
        AQS2      & \ColorCell{0.077} & \ColorCell{0.044} & \ColorCell{0.085} & \ColorCell{0.049} & \ColorCell{0.136} & \ColorCell{0.152} & \ColorCell{0.151} & \ColorCell{0.147} \\
        AQS2DH    & \ColorCell{0.097} & \ColorCell{0.065} & \ColorCell{0.106} & \ColorCell{0.026} & \ColorCell{0.139} & \ColorCell{0.129} & \ColorCell{0.140} & \ColorCell{0.121} \\
        AQS2NBr   & \ColorCell{0.001} & \ColorCell{0.045} & \ColorCell{0.018} & \ColorCell{0.112} & \ColorCell{0.273} & \ColorCell{0.265} & \ColorCell{0.256} & \ColorCell{0.262} \\
        AQDH45CA  & \ColorCell{0.192} & \ColorCell{0.162} & \ColorCell{0.197} & \ColorCell{0.065} & \ColorCell{0.034} & \ColorCell{0.022} & \ColorCell{0.020} & \ColorCell{0.021} \\
        AQDH18MH  & \ColorCell{0.076} & \ColorCell{0.080} & \ColorCell{0.112} & \ColorCell{0.019} & \ColorCell{0.118} & \ColorCell{0.115} & \ColorCell{0.117} & \ColorCell{0.116} \\
        AQTrHM    & \ColorCell{0.068} & \ColorCell{0.079} & \ColorCell{0.113} & \ColorCell{0.020} & \ColorCell{0.127} & \ColorCell{0.128} & \ColorCell{0.126} & \ColorCell{0.125} \\
        AQTH12    & \ColorCell{0.093} & \ColorCell{0.105} & \ColorCell{0.145} & \ColorCell{0.013} & \ColorCell{0.083} & \ColorCell{0.071} & \ColorCell{0.073} & \ColorCell{0.075} \\
        AQTH14    & \ColorCell{0.136} & \ColorCell{0.133} & \ColorCell{0.095} & \ColorCell{0.227} & \ColorCell{0.324} & \ColorCell{0.313} & \ColorCell{0.315} & \ColorCell{0.319} \\
        NQ12S     & \ColorCell{0.159} & \ColorCell{0.156} & \ColorCell{0.220} & \ColorCell{0.082} & \ColorCell{0.002} & \ColorCell{0.020} & \ColorCell{0.026} & \ColorCell{0.023} \\
        NQ14HB    & \ColorCell{0.045} & \ColorCell{0.034} & \ColorCell{0.101} & \ColorCell{0.032} & \ColorCell{0.142} & \ColorCell{0.138} & \ColorCell{0.134} & \ColorCell{0.135} \\
        NQ14H     & \ColorCell{0.174} & \ColorCell{0.160} & \ColorCell{0.230} & \ColorCell{0.091} & \ColorCell{0.001} & \ColorCell{0.001} & \ColorCell{0.003} & \ColorCell{0.000} \\
        BQ14S     & \ColorCell{0.229} & \ColorCell{0.217} & \ColorCell{0.297} & \ColorCell{0.156} & \ColorCell{0.073} & \ColorCell{0.045} & \ColorCell{0.049} & \ColorCell{0.047} \\
        BQ12      & \ColorCell{0.174} & \ColorCell{0.168} & \ColorCell{0.238} & \ColorCell{0.100} & \ColorCell{0.002} & \ColorCell{0.001} & \ColorCell{0.004} & \ColorCell{0.005} \\
        BQ14      & \ColorCell{0.178} & \ColorCell{0.167} & \ColorCell{0.245} & \ColorCell{0.109} & \ColorCell{0.000} & \ColorCell{0.002} & \ColorCell{0.005} & \ColorCell{0.003} \\
        BQ12DS    & \ColorCell{0.115} & \ColorCell{0.128} & \ColorCell{0.199} & \ColorCell{0.060} & \ColorCell{0.039} & \ColorCell{0.063} & \ColorCell{0.046} & \ColorCell{0.056} \\
        BQ14DH    & \ColorCell{0.136} & \ColorCell{0.155} & \ColorCell{0.235} & \ColorCell{0.093} & \ColorCell{0.008} & \ColorCell{0.005} & \ColorCell{0.003} & \ColorCell{0.006} \\
        BQ14DHDCl & \ColorCell{0.121} & \ColorCell{0.144} & \ColorCell{0.228} & \ColorCell{0.080} & \ColorCell{0.020} & \ColorCell{0.019} & \ColorCell{0.019} & \ColorCell{0.020} \\
        BQ14TCl   & \ColorCell{0.123} & \ColorCell{0.102} & \ColorCell{0.213} & \ColorCell{0.066} & \ColorCell{0.034} & \ColorCell{0.035} & \ColorCell{0.041} & \ColorCell{0.043} \\
        BQ14TH    & \ColorCell{0.188} & \ColorCell{0.236} & \ColorCell{0.307} & \ColorCell{0.177} & \ColorCell{0.069} & \ColorCell{0.069} & \ColorCell{0.074} & \ColorCell{0.075} \\
        BQ14TF    & \ColorCell{0.101} & \ColorCell{0.095} & \ColorCell{0.181} & \ColorCell{0.039} & \ColorCell{0.064} & \ColorCell{0.063} & \ColorCell{0.066} & \ColorCell{0.067} \\
        \midrule
        MAE       & \ColorCell{0.129} & \ColorCell{0.126} & \ColorCell{0.168} & \ColorCell{0.083} & \ColorCell{0.099} & \ColorCell{0.097} & \ColorCell{0.097} & \ColorCell{0.096} \\
        \bottomrule
    \end{tabular}
    \caption{Absolute errors of PCET redox potential for different methods on Test Set B (unit in V). For DFT methods, geometries were optimized using the def2-SVPD basis set, followed by single-point energy calculations with the def2-TZVPD basis set.}
    \label{tab:pcet_ph}
\end{table*}

\begin{table*}[tbp]
    \centering
    \begin{tabular}{lcccccccc}
        \toprule
        OPT Method & B3LYP-D3(BJ) & M06-2X-D3 & $\omega$B97X-D3(BJ) & $\omega$B97M-D3(BJ) & \multicolumn{2}{c}{MACE-OMol} & \multicolumn{2}{c}{UMA-s} \\
        \midrule
        SP Method & B3LYP-D3(BJ) & M06-2X-D3 & $\omega$B97X-D3(BJ) & $\omega$B97M-D3(BJ) & MACE-OMol & $\omega$B97M-V & UMA-s & $\omega$B97M-V \\
        \midrule
        BQ14*  & \ColorCell{0.000} & \ColorCell{0.000} & \ColorCell{0.000} & \ColorCell{0.000} & \ColorCell{0.000} & \ColorCell{0.000} & \ColorCell{0.000} & \ColorCell{0.000} \\
        BQ14Ph & \ColorCell{0.049} & \ColorCell{0.029} & \ColorCell{0.041} & \ColorCell{0.040} & \ColorCell{0.014} & \ColorCell{0.052} & \ColorCell{0.069} & \ColorCell{0.027} \\
        BQ14Me & \ColorCell{0.021} & \ColorCell{0.013} & \ColorCell{0.014} & \ColorCell{0.013} & \ColorCell{0.013} & \ColorCell{0.008} & \ColorCell{0.017} & \ColorCell{0.004} \\
        BQ14tBu & \ColorCell{0.042} & \ColorCell{0.026} & \ColorCell{0.038} & \ColorCell{0.031} & \ColorCell{0.035} & \ColorCell{0.017} & \ColorCell{0.033} & \ColorCell{0.010} \\
        BQ14MeO & \ColorCell{0.085} & \ColorCell{0.063} & \ColorCell{0.065} & \ColorCell{0.063} & \ColorCell{0.031} & \ColorCell{0.076} & \ColorCell{0.081} & \ColorCell{0.062} \\
        BQ14DMe26 & \ColorCell{0.047} & \ColorCell{0.035} & \ColorCell{0.035} & \ColorCell{0.032} & \ColorCell{0.008} & \ColorCell{0.027} & \ColorCell{0.033} & \ColorCell{0.021} \\
        BQ14DMe23 & \ColorCell{0.017} & \ColorCell{0.035} & \ColorCell{0.022} & \ColorCell{0.020} & \ColorCell{0.007} & \ColorCell{0.008} & \ColorCell{0.033} & \ColorCell{0.011} \\
        BQ14TrMe & \ColorCell{0.037} & \ColorCell{0.055} & \ColorCell{0.036} & \ColorCell{0.033} & \ColorCell{0.008} & \ColorCell{0.024} & \ColorCell{0.037} & \ColorCell{0.020} \\
        BQ14DMeO26 & \ColorCell{0.136} & \ColorCell{0.101} & \ColorCell{0.102} & \ColorCell{0.099} & \ColorCell{0.036} & \ColorCell{0.115} & \ColorCell{0.111} & \ColorCell{0.102} \\
        BQ14TMe & \ColorCell{0.045} & \ColorCell{0.086} & \ColorCell{0.051} & \ColorCell{0.044} & \ColorCell{0.096} & \ColorCell{0.105} & \ColorCell{0.082} & \ColorCell{0.066} \\
        DDQ & \ColorCell{0.067} & \ColorCell{0.114} & \ColorCell{0.123} & \ColorCell{0.087} & \ColorCell{0.065} & \ColorCell{0.064} & \ColorCell{0.059} & \ColorCell{0.109} \\
        BQ12TF & \ColorCell{0.029} & \ColorCell{0.060} & \ColorCell{0.058} & \ColorCell{0.036} & \ColorCell{0.321} & \ColorCell{0.016} & \ColorCell{0.049} & \ColorCell{0.044} \\
        BQ14DCl25 & \ColorCell{0.037} & \ColorCell{0.013} & \ColorCell{0.011} & \ColorCell{0.026} & \ColorCell{0.047} & \ColorCell{0.031} & \ColorCell{0.095} & \ColorCell{0.009} \\
        BQ14TCl & \ColorCell{0.033} & \ColorCell{0.020} & \ColorCell{0.024} & \ColorCell{0.005} & \ColorCell{0.047} & \ColorCell{0.007} & \ColorCell{0.086} & \ColorCell{0.017} \\
        BQ14Cl & \ColorCell{0.015} & \ColorCell{0.037} & \ColorCell{0.003} & \ColorCell{0.010} & \ColorCell{0.020} & \ColorCell{0.017} & \ColorCell{0.118} & \ColorCell{0.006} \\
        NQ14 & \ColorCell{0.043} & \ColorCell{0.042} & \ColorCell{0.065} & \ColorCell{0.059} & \ColorCell{0.006} & \ColorCell{0.049} & \ColorCell{0.016} & \ColorCell{0.032} \\
        AQ & \ColorCell{0.090} & \ColorCell{0.098} & \ColorCell{0.154} & \ColorCell{0.143} & \ColorCell{0.169} & \ColorCell{0.151} & \ColorCell{0.075} & \ColorCell{0.103} \\
        AQDCl18 & \ColorCell{0.280} & \ColorCell{0.316} & \ColorCell{0.357} & \ColorCell{0.351} & \ColorCell{0.338} & \ColorCell{0.371} & \ColorCell{0.352} & \ColorCell{0.327} \\
        NQ14DCl23 & \ColorCell{0.014} & \ColorCell{0.042} & \ColorCell{0.027} & \ColorCell{0.017} & \ColorCell{0.094} & \ColorCell{0.034} & \ColorCell{0.027} & \ColorCell{0.045} \\
        BQ12DtBu35 & \ColorCell{0.140} & \ColorCell{0.111} & \ColorCell{0.146} & \ColorCell{0.131} & \ColorCell{0.254} & \ColorCell{0.143} & \ColorCell{0.101} & \ColorCell{0.103} \\
        BQ12tBu4 & \ColorCell{0.083} & \ColorCell{0.066} & \ColorCell{0.087} & \ColorCell{0.080} & \ColorCell{0.287} & \ColorCell{0.104} & \ColorCell{0.046} & \ColorCell{0.063} \\
        PQ & \ColorCell{0.111} & \ColorCell{0.114} & \ColorCell{0.185} & \ColorCell{0.168} & \ColorCell{0.070} & \ColorCell{0.190} & \ColorCell{0.098} & \ColorCell{0.142} \\
        NQ12 & \ColorCell{0.072} & \ColorCell{0.069} & \ColorCell{0.104} & \ColorCell{0.097} & \ColorCell{0.212} & \ColorCell{0.096} & \ColorCell{0.036} & \ColorCell{0.082} \\
        Phendio & \ColorCell{0.085} & \ColorCell{0.101} & \ColorCell{0.173} & \ColorCell{0.165} & \ColorCell{0.086} & \ColorCell{0.168} & \ColorCell{0.170} & \ColorCell{0.132} \\
        BQ12TCl & \ColorCell{0.000} & \ColorCell{0.054} & \ColorCell{0.046} & \ColorCell{0.022} & \ColorCell{0.236} & \ColorCell{0.015} & \ColorCell{0.058} & \ColorCell{0.031} \\
        \midrule
        MAE & \ColorCell{0.063} & \ColorCell{0.068} & \ColorCell{0.079} & \ColorCell{0.071} & \ColorCell{0.100} & \ColorCell{0.076} & \ColorCell{0.075} & \ColorCell{0.063} \\
        \bottomrule
    \end{tabular}
    \caption{Absolute errors of \ce{1e-} ET redox potential for different methods on Test Set C (unit in V). For DFT methods, geometries were optimized using the def2-SVPD basis set, followed by single-point energy calculations with the def2-TZVPD basis set. * Compounds marked as the reference.}
    \label{tab:et_jacs}
\end{table*}

\begin{table*}[tbp]
    \centering
    \begin{tabular}{lcccccccc}
        \toprule
        OPT Method & B3LYP-D3(BJ) & M06-2X-D3 & $\omega$B97X-D3(BJ) & $\omega$B97M-D3(BJ) & \multicolumn{2}{c}{MACE-OMol} & \multicolumn{2}{c}{UMA-s} \\
        \midrule
        SP Method & B3LYP-D3(BJ) & M06-2X-D3 & $\omega$B97X-D3(BJ) & $\omega$B97M-D3(BJ) & MACE-OMol & $\omega$B97M-V & UMA-s & $\omega$B97M-V \\
        \midrule
        BQ14*  & \ColorCell{0.047} & \ColorCell{0.047} & \ColorCell{0.047} & \ColorCell{0.047} & \ColorCell{0.047} & \ColorCell{0.047} & \ColorCell{0.047} & \ColorCell{0.047} \\
        BQ14Ph & \ColorCell{0.015} & \ColorCell{0.020} & \ColorCell{0.025} & \ColorCell{0.025} & \ColorCell{0.028} & \ColorCell{0.026} & \ColorCell{0.028} & \ColorCell{0.026} \\
        BQ14Me & \ColorCell{0.023} & \ColorCell{0.023} & \ColorCell{0.026} & \ColorCell{0.026} & \ColorCell{0.029} & \ColorCell{0.029} & \ColorCell{0.028} & \ColorCell{0.028} \\
        BQ14tBu & \ColorCell{0.023} & \ColorCell{0.022} & \ColorCell{0.025} & \ColorCell{0.026} & \ColorCell{0.030} & \ColorCell{0.028} & \ColorCell{0.027} & \ColorCell{0.027} \\
        BQ14MeO & \ColorCell{0.013} & \ColorCell{0.008} & \ColorCell{0.005} & \ColorCell{0.005} & \ColorCell{0.002} & \ColorCell{0.002} & \ColorCell{0.002} & \ColorCell{0.002} \\
        BQ14DMe26 & \ColorCell{0.029} & \ColorCell{0.028} & \ColorCell{0.034} & \ColorCell{0.035} & \ColorCell{0.033} & \ColorCell{0.033} & \ColorCell{0.032} & \ColorCell{0.032} \\
        BQ14DMe23 & \ColorCell{0.018} & \ColorCell{0.012} & \ColorCell{0.026} & \ColorCell{0.027} & \ColorCell{0.014} & \ColorCell{0.015} & \ColorCell{0.017} & \ColorCell{0.017} \\
        BQ14TrMe & \ColorCell{0.018} & \ColorCell{0.008} & \ColorCell{0.027} & \ColorCell{0.028} & \ColorCell{0.011} & \ColorCell{0.012} & \ColorCell{0.016} & \ColorCell{0.017} \\
        BQ14DMeO26 & \ColorCell{0.005} & \ColorCell{0.013} & \ColorCell{0.021} & \ColorCell{0.019} & \ColorCell{0.020} & \ColorCell{0.022} & \ColorCell{0.021} & \ColorCell{0.022} \\
        BQ14TMe & \ColorCell{0.022} & \ColorCell{0.006} & \ColorCell{0.037} & \ColorCell{0.041} & \ColorCell{0.024} & \ColorCell{0.029} & \ColorCell{0.004} & \ColorCell{0.005} \\
        DDQ & \ColorCell{0.027} & \ColorCell{0.017} & \ColorCell{0.010} & \ColorCell{0.024} & \ColorCell{0.014} & \ColorCell{0.017} & \ColorCell{0.017} & \ColorCell{0.019} \\
        BQ12TF & \ColorCell{0.054} & \ColorCell{0.048} & \ColorCell{0.041} & \ColorCell{0.048} & \ColorCell{0.041} & \ColorCell{0.043} & \ColorCell{0.046} & \ColorCell{0.045} \\
        BQ14DCl25 & \ColorCell{0.002} & \ColorCell{0.007} & \ColorCell{0.016} & \ColorCell{0.010} & \ColorCell{0.016} & \ColorCell{0.013} & \ColorCell{0.010} & \ColorCell{0.011} \\
        BQ14TCl & \ColorCell{0.033} & \ColorCell{0.026} & \ColorCell{0.010} & \ColorCell{0.023} & \ColorCell{0.012} & \ColorCell{0.015} & \ColorCell{0.017} & \ColorCell{0.017} \\
        BQ14Cl & \ColorCell{0.016} & \ColorCell{0.038} & \ColorCell{0.023} & \ColorCell{0.020} & \ColorCell{0.025} & \ColorCell{0.022} & \ColorCell{0.020} & \ColorCell{0.021} \\
        NQ14 & \ColorCell{0.043} & \ColorCell{0.034} & \ColorCell{0.023} & \ColorCell{0.023} & \ColorCell{0.035} & \ColorCell{0.035} & \ColorCell{0.038} & \ColorCell{0.037} \\
        AQ & \ColorCell{0.004} & \ColorCell{0.043} & \ColorCell{0.074} & \ColorCell{0.073} & \ColorCell{0.052} & \ColorCell{0.047} & \ColorCell{0.058} & \ColorCell{0.053} \\
        AQDCl18 & \ColorCell{0.077} & \ColorCell{0.099} & \ColorCell{0.141} & \ColorCell{0.144} & \ColorCell{0.149} & \ColorCell{0.146} & \ColorCell{0.152} & \ColorCell{0.144} \\
        NQ14DCl23 & \ColorCell{0.046} & \ColorCell{0.052} & \ColorCell{0.050} & \ColorCell{0.059} & \ColorCell{0.046} & \ColorCell{0.047} & \ColorCell{0.048} & \ColorCell{0.048} \\
        BQ12DtBu35 & \ColorCell{0.066} & \ColorCell{0.068} & \ColorCell{0.063} & \ColorCell{0.063} & \ColorCell{0.075} & \ColorCell{0.074} & \ColorCell{0.070} & \ColorCell{0.073} \\
        BQ12tBu4 & \ColorCell{0.025} & \ColorCell{0.029} & \ColorCell{0.024} & \ColorCell{0.023} & \ColorCell{0.032} & \ColorCell{0.030} & \ColorCell{0.026} & \ColorCell{0.029} \\
        PQ & \ColorCell{0.003} & \ColorCell{0.018} & \ColorCell{0.047} & \ColorCell{0.045} & \ColorCell{0.061} & \ColorCell{0.059} & \ColorCell{0.044} & \ColorCell{0.045} \\
        NQ12 & \ColorCell{0.001} & \ColorCell{0.009} & \ColorCell{0.019} & \ColorCell{0.021} & \ColorCell{0.009} & \ColorCell{0.007} & \ColorCell{0.011} & \ColorCell{0.010} \\
        Phendio & \ColorCell{0.105} & \ColorCell{0.151} & \ColorCell{0.180} & \ColorCell{0.179} & \ColorCell{0.180} & \ColorCell{0.175} & \ColorCell{0.178} & \ColorCell{0.178} \\
        BQ12TCl & \ColorCell{0.007} & \ColorCell{0.023} & \ColorCell{0.024} & \ColorCell{0.016} & \ColorCell{0.025} & \ColorCell{0.026} & \ColorCell{0.022} & \ColorCell{0.022} \\
        \midrule
        MAE & \ColorCell{0.029} & \ColorCell{0.034} & \ColorCell{0.041} & \ColorCell{0.042} & \ColorCell{0.040} & \ColorCell{0.040} & \ColorCell{0.039} & \ColorCell{0.039} \\
        \bottomrule
    \end{tabular}
    \caption{Absolute errors of PCET redox potential for different methods on Test Set C (unit in V). For DFT methods, geometries were optimized using the def2-SVPD basis set, followed by single-point energy calculations with the def2-TZVPD basis set. * Compounds marked as the reference.}
    \label{tab:pcet_jacs}
\end{table*}

\section{Results} 
There are two primary types of redox reactions involving molecules in flow-based electrochemistry: ET and PCET processes in aqueous solutions. 
To benchmark these two types of reactions, we selected three representative studies with experimentally reported redox potentials. Figure \ref{fig:workflow}c shows the characteristic reactions and molecules in the three groups,
including ET in Lewis base molecules \cite{li_lewis-base_2022} (blue panel, Test Set A), PCET at pH=0 for quinones (mainly functionalized by polar groups, green panel, Test Set B) \cite{fornari_pcet-ph_2021}, and ET/PCET for quinones (mainly functionalized by non-polar groups, orange panel, Test Set C) \cite{huynh_pcet-et_2016}.

\subsection{ET reactions of Lewis bases}

\citet{li_lewis-base_2022} designed redox-tunable Lewis bases for reversible \ce{CO2} capture in organic solvent systems by reducing or oxidizing these \ce{sp^2}-nitrogen-centered Lewis bases. As the redox potential is critical to this tunability, we used the reported experimental results in Ref.~\cite{li_lewis-base_2022} for the benchmark test. Specifically, the solvent employed in this study is dimethyl sulfoxide (DMSO). The reference electrode is the ferrocenium/ferrocene (\ce{Fc^+/Fc}) couple, which has a reference potential of 4.84 V, resulting from \ce{Fc^+/Fc} relative to standard hydrogen electrode (SHE) at 0.40 V \cite{gagne_ferrocene_1980}, and SHE relative to vacuum level at 4.44 V \cite{trasatti_she_1986}.

Table~\ref{tab:lewis} displays the absolute errors associated with the redox potential of a series of DFT functionals and MACE-OMol for the Lewis bases. 
The B3LYP-D3(BJ) exhibits the highest mean absolute error (MAE) of 0.173 V, while the M06-2X-D3 method exhibits the second-highest MAE of 0.130 V. 
In contrast, range-separated functionals $\omega$B97X-D3BJ and $\omega$B97M-D3BJ demonstrate better overall performance (MAE of 0.093 V and 0.096 V, respectively). 
Range-separated functionals satisfy the correct asymptotic behavior of the exchange potential and significantly reduce self-interaction error, collectively leading to their enhanced accuracy across various computed properties \cite{bremond_rangeseparated_2018}.

FPs perform differently for the \ce{1e-} ET redox reactions. UMA-s gives a low MAE of 0.050 V for the \ce{1e-} ET redox potential, achieving accuracy comparable to the best DFT methods as tested. Yet the MACE-OMol performs reasonably well (MAE: 0.162 V). Both FPs show significant errors when predicting the \ce{2e-} ET (MAE: 1.925 V for MACE-OMol and MAE: 1.449 V for UMA-s).

We performed the single-point DFT calculations for correction at the target level of theory ($\omega$B97M-V/def2-TZVPD) with the FP-optimized structure. 
Notably, the single-point correction substantially reduces redox potential prediction error for \ce{1e-} ET for MACE-OMol (MAE from 0.162 to 0.029 V). The correction for UMA-s maintains high accuracy (MAE from 0.050 V to 0.037 V).
Conversely, the error of \ce{2e-} ET remains relatively large (MAE: 0.279 V and 0.226 V for MACE-OMol and UMA-s, respectively). 
This suggests that while FPs can fail to accurately predict the \ce{1e-} ET reactive ion energies, they nonetheless yield reliable predictions for equilibrium configurations and vibrational frequencies.
To confirm this, 2,1,3-benzothiadiazole (BNSN) is used as a case study: we optimized the \ce{1e-} and \ce{2e-} ET product structures and calculated Hessian matrices using both FPs and the target DFT (Figure \ref{fig:hess}). For the \ce{1e-} ET product, the Hessian error is small (MAE: 0.089 $\text{eV}/\text{\AA}^2$ for MACE-OMol and 0.105 $\text{eV}/\text{\AA}^2$ for UMA-s, respectively), whereas the \ce{2e-} ET product error is much higher (MAE: 0.74 $\text{eV}/\text{\AA}^2$ 0.736 for MACE-OMol and $\text{eV}/\text{\AA}^2$ for UMA-s, respectively).

The comparison indicates that, for \ce{1e-} ET species, both FPs reasonably reproduce the gradients and Hessians that guide the optimization toward the optimal ground-state conformation.Consequently, the free energy corrections derived from these gradients/Hessians retain their accuracy when combined with DFT single-point corrections. 
Conversely, the high Hessian errors associated with the \ce{2e-} ET product indicate a heightened inconsistency in the predicted equilibrium conformation and vibrational properties. This discrepancy leads to erroneous predictions when the charge or spin multiplicity becomes more extreme.

\subsection{PCET reactions of quinones with polar groups}

The second type of reaction relevant to flow-based electrochemical carbon removal is the PCET reaction, i.e., \ce{Q + 2H+ + 2e-} $\rightarrow$ \ce{H2Q}. We adopted the experimentally reported PCET redox potentials from Ref.~\cite{fornari_pcet-ph_2021} and \cite{wedege_ph-exptl_2016} for the second group of benchmarks. 
The test set includes 28 quinone compounds, some bearing polar functional groups (e.g., sulfonic acid, amino, and hydroxyl groups) that impart excellent water solubility. 
Given that the PCET reaction involves protons, we followed the original protocol and employed the reported absolute Gibbs free energy of the aqueous proton, $G_{(\text{aq})} \left(\ce{H+}\right) = -11.45$ eV~\cite{tissandier_protons_1998,zhan_protons_2001}. 
The experimental data report potentials relative to the SHE in the aqueous phase, with the SHE referenced to the vacuum level at 4.44 V \cite{trasatti_she_1986}.

The benchmark results are presented in Table~\ref{tab:pcet_ph}. DFT calculations follow a similar trend to the previous results: range-separated functionals performed better overall, with $\omega$B97M-D3(BJ) yielding the best results (MAE: 0.083 V). In addition, MACE-OMol and UMA-s perform satisfactorily on this dataset, achieving MAEs of 0.099 V and 0.097 V. Notably, the subsequent target DFT single-point energy correction resulted in a marginal change to the MAE (0.097 V and 0.096 V, respectively), suggesting that FPs already provide highly accurate redox potentials for these PCET reactions.
Unlike its poor performance with reactive ions forming via ET, FPs show greater precision in predicting the energies of neutral molecules or those with ionic functional groups (e.g., \ce{SO3-}), consistently demonstrating comparable results to the $\omega$B97M-V DFT.

Conversely, although PFs and other higher-level DFT methods generally perform well, for molecules such as AQS2NBr and AQTH14, MACE-OMol and UMA-s consistently yield higher redox potential errors than lower-level DFT functionals. We validated single-point energy calculations against reference results from the coupled-cluster method, DLPNO-CCSD(T)-F12 \cite{cizek_cc_1966,purvis_cc_1982,raghavachari_cc_1989,riplinger_dlpno_2013,riplinger_dlpno_2016,saitow_dlpno_2017,guo_dlpno_2018,adler_f12_2007} (see SI Table II to V). Notably, FPs and their target DFT methods show poor consistency with the coupled-cluster method, indicating limitations of target DFT itself for some systems. However, even when using the single-point energy of the coupling cluster, there remains a discrepancy with the experimental results (errors of 0.1 V).
These elevated errors are therefore likely attributable to approximations inherent in the implicit solvent model. As noted in Ref.~\cite{fornari_pcet-ph_2021}, the error associated with AQTH14 decreases when explicit solvent molecules are included in calculations.

\subsection{ET/PCET of quinone-based molecules}

\citet{huynh_pcet-et_2016} identified systematic scaling relationships for ET and PCET in quinones via combined experimental and DFT studies. The quinones in their study feature predominantly nonpolar substituents (e.g., alkyl, alkoxy, halogen groups). 
These differ from the polar-substituted quinones in the previous test set \cite{fornari_pcet-ph_2021}, making them well-suited as complementary systems for benchmarking the quinone redox potentials.

We used the benzoquinone (BQ) as the reference to derive the PCET redox potentials in the Test Set C.
Specifically, the \ce{1e-} ET and PCET redox potentials of BQ are fixed at -0.8815 V and 0.690 V as the reference potentials $\mathcal{E}^\circ_\text{ref}$ in Ref.~\cite{huynh_pcet-et_2016}.
A shifted term $\Delta \mathcal{E} = \mathcal{E}^\circ_\text{ref} \left( \text{BQ} \right) - \mathcal{E}^\circ_\text{calc} \left( \text{BQ} \right)$ was applied to the calculated redox potentials $\mathcal{E}^\circ_\text{calc} \left(\cdot\right)$ of other target reactions, ensuring these values align with the BQ reference (see SI for details).
The MAEs of ET and PCET are shown in Table~\ref{tab:et_jacs} and \ref{tab:pcet_jacs}. A notable observation is that calibrating calculations using the experimental redox potential of BQ substantially mitigates errors arising from systematic shifts. The MAEs are lower than 0.1 V for all tested methods. 
Specifically, for PCET reactions, the MAEs are further reduced to approximately 0.04 V (see Table~\ref{tab:pcet_jacs}). The discrepancies in performance between different DFT methods are markedly diminished by this calibration approach, indicating the consistency of relative trends across distinct redox pairs. 

Consistent with FPs' previous performances on ET and PCET reactions, MACE-OMol struggles to accurately predict the energies of ions with a transferred electron (MAE 0.100 V), leading to large errors for outliers (particularly, for BQ12TF, BQ12DtBu35, and BQ12TCl in Table~\ref{tab:et_jacs}). In contrast, UMA-s demonstrates a notably strong intrinsic performance on this \ce{1e-} ET test set, achieving a superior MAE of 0.075 V.

As demonstrated in our suggested workflow, this limitation of MACE-OMol can be mitigated by a single-point correction with the target DFT, reducing the MAE from 0.100 V to 0.076 V. Similarly, the hybrid approach can further enhance the accuracy for UMA-s, reducing the MAE from 0.075 V to 0.063 V, which yields the lowest error among all tested methods.

For PCET reactions that do not involve anionic radicals, both FPs perform impressively well. As shown in the last two columns in Table~\ref{tab:pcet_jacs}, the direct MACE-OMol and UMA-s calculations yield MAEs of 0.040 V and 0.039 V in redox potential prediction, respectively, which exactly match their target DFT results.
In contrast to the challenges encountered in multi-ET reactions, the enhanced performance of FPs suggests their accuracy depends on the chemical species involved, excelling for charge-neutral molecules but struggling with underrepresented reactive anions. 

\subsection{Computational efficiency of Hessians}

Computing the Hessian matrix for the optimized structure is often the most computationally intensive step. FPs enable efficient Hessian matrix calculations than DFT. To demonstrate the acceleration in the hybrid workflow, we selected BNSN as a case study to compare computational time between two DFT methods ($\omega$B97M-D3BJ/def2-SVPD and $\omega$B97M-V/def2-TZVPD) and the FPs (MACE-OMol and UMA-s).

\begin{figure}[tbp]
    \centering
    \includegraphics[width=\linewidth]{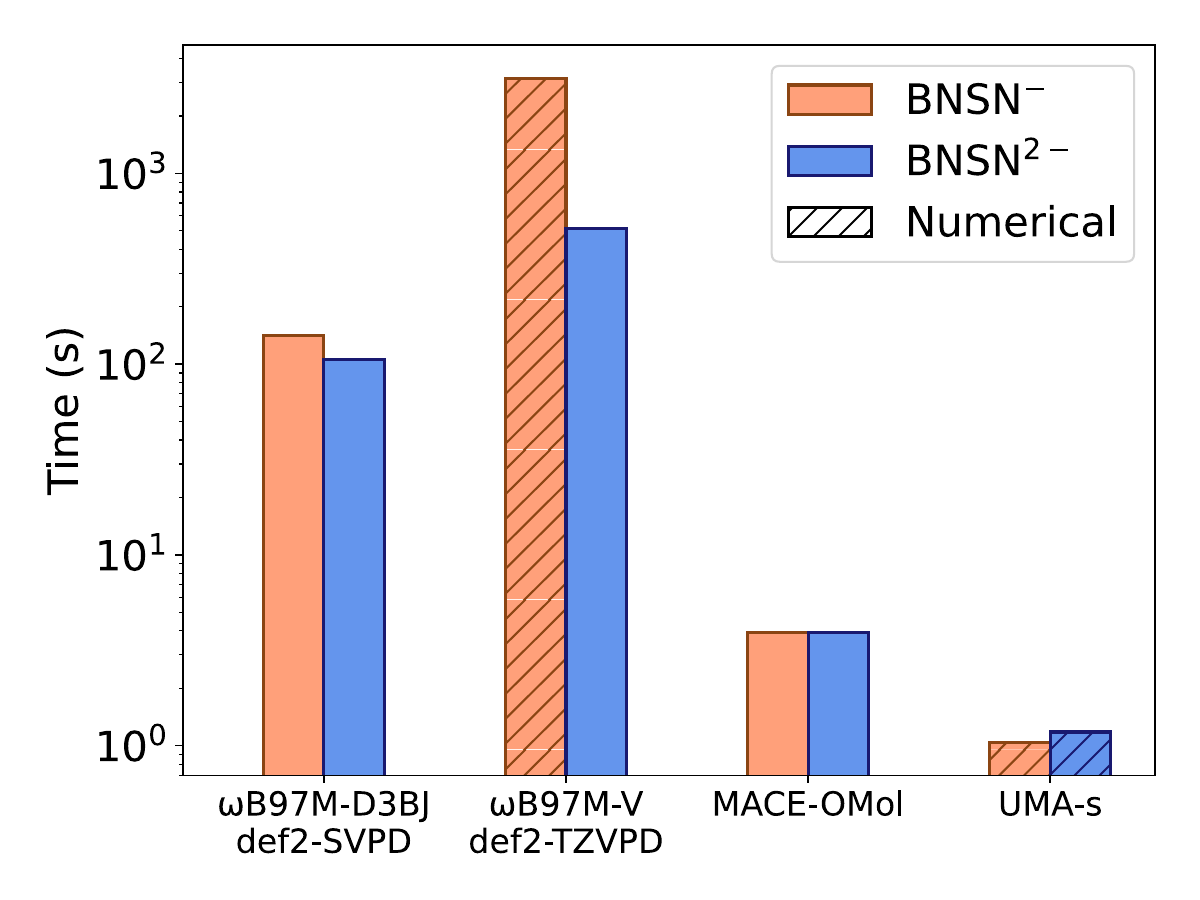}
    \caption{Computational time required for Hessian matrix calculations of open-shell \ce{BNSN-} (orange) and closed-shell \ce{BNSN^2-} (blue) using DFT and FPs. Dashed bars indicate the use of numerical Hessian calculations.
    }
    \label{fig:time}
\end{figure}

We computed the Hessian matrices for the open-shell \ce{BNSN-} and closed-shell \ce{BNSN^2-} anions using all four methods on NVIDIA A40 GPUs, with the corresponding computational timings summarized in Figure~\ref{fig:time}. Notably, the VV10 correlation in PySCF lacks support for Hessian calculations on open-shell systems, and the UMA-s model does not provide analytical Hessian evaluations. 
To accommodate these limitations, numerical finite-difference gradients were used for the affected cases. 
For the 13-atom systems, FPs yield Hessian matrices in only a few seconds, while DFT with analytical Hessians requires hundreds of seconds, and numerical Hessian requires over 3000 seconds. This shows the speed advantage of FPs for Hessian calculations, while maintaining comparable accuracy to DFT with the hybrid workflow.

\section{Discussion}

High-throughput screening with quantum chemistry and foundation potentials (FP) has emerged as an indispensable tool for the discovery of functional molecules and materials \cite{batatia_mace_2023, levine_omol25_2025}. 
In the context of electrochemical carbon removal, estimating redox potentials by evaluating Gibbs free energies at various charge/spin states and solvent conditions is crucial for screening redox-active molecules, not only for electrochemically induced carbon capture  \cite{huynh_pcet-et_2016, li_lewis-base_2022}, but also for energy storage applications \cite{huskinson_metal-free_2014, liang_universal_2017} and other redox-mediated systems \cite{zhang_decoupled_2021, wang_materials_2025}.
While DFT enables accurate quantum-mechanical calculations, its computational cost remains a barrier to large-scale materials screening. FPs trained on extensive DFT datasets provide a promising alternative for efficient evaluations.
A key question remains: Can as-pretrained FPs be used reliably for such high-throughput screening?

\begin{figure*}[tbp]
    \centering
    \includegraphics[width=\linewidth]{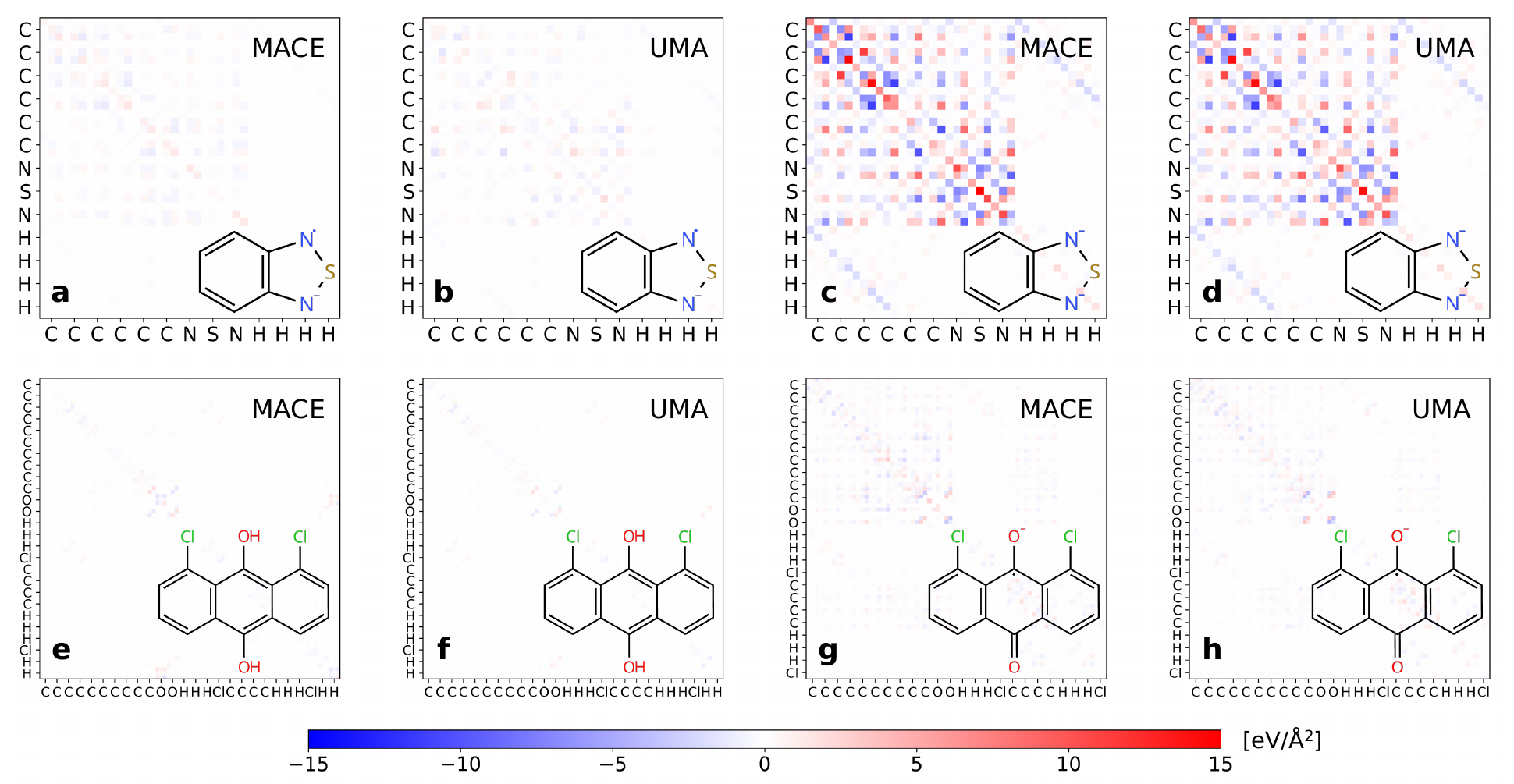}
    \caption{
    Errors in Hessian matrices calculated by MACE-OMol and UMA-s relative to the target DFT ($\omega$B97M-V) reference. Panels show results for: (a, b) \ce{BNSN-} (ET \ce{1e-}); (c, d) \ce{BNSN^2-} (ET \ce{2e-}); (e, f) \ce{H2AQDCl} (PCET); and (g, h) \ce{AQDCl-} (ET \ce{1e-}). 
    } 
    \label{fig:hess}
\end{figure*}

In this study, we compared the redox potential predictions for various ET and PCET reactions using different levels of theory as well as  MACE-OMol and UMA-s FP. 
Our results show that the two FPs perform exceptionally well in predicting PCET redox potentials, consistently yielding high agreement with experimentally reported redox potentials, as well as achieving accuracy comparable to target DFT calculations not only for single-point energies but also for gradients such as Hessian matrices (see Figure~\ref{fig:hess}c).  
This translates to reliable predictions of equilibrium structures and thermodynamic properties. Notably, despite lacking direct supervision on Hessian matrices, FPs still predict them effectively by learning from energies and forces.

However, their performance on ET-derived ions reveals a differential in intrinsic accuracy. UMA-s demonstrates superior performance for \ce{1e-} ET compared to MACE-OMol (Table~\ref{tab:lewis} and Table~\ref{tab:et_jacs}). The DFT single-point correction is essential to mitigate MACE-OMol's higher intrinsic energy error, reducing its \ce{1e-} ET MAE substantially.
We calculated the Hessian matrix for the FP-optimized geometry using both the FP and DFT ($\omega$B97M-V). The difference is shown in Figure~\ref{fig:hess}a, where the FP results are nearly identical to those provided by DFT. 
Therefore, the DFT can be used directly to refine the single-electron energies to achieve good agreement with the experimental redox potential. Despite the effectiveness of the hybrid workflow for \ce{1e-} ET, both FPs struggle significantly with \ce{2e-} ET processes. For instance, the predictions of MACE-OMol and UMA-s for the \ce{2e-} reduction of BNSN exhibit substantial deviations from experimental data, with the errors of 0.237 V and 0.195 V for the resulting \ce{BNSN^2-} species, respectively.

We found that the Hessian matrices reveal significant discrepancies between DFT and FPs (Figure \ref{fig:hess}b), indicating that the optimized conformation deviates from the ground state.
This discrepancy arises from a failure mode analogous to ``hallucination", a known challenge for large ML models trained with supervised learning \cite{kalai_why_2025}. Such models can produce physically unreliable or nonsensical predictions when operating on out-of-distribution data. 
This problem is particularly characteristic of architectures that embed discrete chemical states (e.g., charge and spin) as one-hot-encoded features. The model's ability to accurately interpret these features relies entirely on extensive supervision from the training data \cite{yuan_foundation_2025}. 
Therefore, while the OMol25 dataset includes a variety of charge and spin states, the construction of its relevant electrolyte subset intentionally focused on sampling systems involving only the gain or loss of a single electron \cite{levine_omol25_2025}, resulting in the significant errors in predicting energies for dianions in \ce{2e-}. This underscores the model's limited transferability to underrepresented chemical environments and confirms that a final DFT correction is essential for achieving reliable predictions.

To demonstrate this ``hallucination", we compared the energy, structure, and Hessian of the \ce{2e-} reduced state (\ce{BNSN^2-}) predicted by FPs against the same properties for the neutral BNSN molecule. The results from the target DFT calculation for \ce{BNSN^2-} served as the reference.

\begin{table}[tb]
    \centering
    \begin{tabular}{lccc}
        \toprule
         & \multirow{2}{*}{Model} & \multicolumn{2}{c}{Compared Target} \\
         & & FP-BNSN & DFT-\ce{BNSN^2-} \\
        \midrule
        \multirow{2}{*}{$\left| \Delta G \right|$ (eV)} & MACE-OMol & 0.25 & 3.40 \\
                                                        & UMA-s     & 0.06 & 3.09 \\
        \multirow{2}{*}{$\left| \Delta E \right|$ (eV)} & MACE-OMol & 0.22 & 3.59 \\
                                                        & UMA-s     & 0.11 & 3.25 \\
        \multirow{2}{*}{RMSD (\AA)}                     & MACE-OMol & 0.002 & 0.050 \\
                                                        & UMA-s     & 0.007 & 0.047 \\
        \multirow{2}{*}{$\mathbf{H}$ MAE (eV/\AA$^2$)}  & MACE-OMol & 0.008 & 0.747 \\
                                                        & UMA-s     & 0.192 & 0.736 \\
        \bottomrule
    \end{tabular}
    \caption{Error comparison of FP-predicted \ce{BNSN^2-} properties against FP-predicted BNSN and DFT-calculated \ce{BNSN^2-}. The properties compared include the absolute errors in free energy, single-point energy, the root mean square deviation, and the MAE of the Hessian.}
    \label{tab:error_compare}
\end{table}

Table~\ref{tab:error_compare} shows that the properties predicted by the FPs for \ce{BNSN^2-} closely resemble those predicted for the neutral BNSN molecule, while deviating significantly from the reference values from DFT ($\omega$B97M-V/def2-TZVPD). The comparison demonstrates that FPs incorrectly map the \ce{BNSN^2-} onto the neutral BNSN configuration, which is likely due to the underrepresentation of \ce{2e-} redox species in the training set. The inconsistent mapping leads to substantial errors in property prediction.

Another practical gap for using FPs in electrochemical redox potential calculations is handling the solvation effect. As MACE-OMol and UMA are pretrained on gas-phase DFT calculations, implicit solvation models cannot be directly applied, since electronic structure information is required but FPs do not provide it. The Born-Haber cycle offers a pragmatic workaround that computes gas-phase free energies using computationally efficient FPs and the solvation free energy using a separate external correction. While models like the Polarizable Continuum Model (PCM) are common \cite{tomasi_pcm_2005}, the SMD is particularly well-suited for FP-based computational workflows. 
There are two primary advantages to using SMD. First, it systematically parameterizes non-electrostatic contributions \cite{marenich_smd_2009}, often leading to more accurate solvation free energies. 
Second, and most crucially, the SMD model was developed with empirical parameters optimized using gas-phase optimized molecular configurations against experimental solvation energies \cite{sola_menshutkin_1991}.
The SMD model ensures direct compatibility with FP-optimized gas-phase structures, enabling seamless integration for redox potential calculations with the Born-Haber cycle.

In summary, this work provides a series of benchmarks of FPs for molecular redox potential calculations compared to several quantum chemistry methods. Our findings demonstrate its exceptional performance for PCET reactions but also reveal inaccuracies for multi-electron transfer processes, a limitation attributed to out-of-distribution predictions for underrepresented charge and spin states.
We therefore propose an optimal computational workflow that leverages the efficiency of FPs for structural optimization and thermochemical corrections, coupled with a necessary single-point energy refinement from DFT and a compatible SMD solvation correction. This pragmatic and hybrid approach represents a more robust and scalable strategy for accelerating the computational discovery of materials for sustainable applications.

\section*{Data Avalability}
The supporting data and codebase are available at
\url{https://github.com/AM3GroupHub/redox_benchmark}.

\section*{Acknowledgements} 
This work was supported by the NUS Presidential Young Professorship startup funding and the Institute of Functional Intelligent Materials (IFIM). 
The computational work was performed on computational resources at the National Supercomputing Center of Singapore (NSCC) and NUS-HPC (CFP03-CF-029).
Y.J. acknowledges the support from the NUS-AISI Joint Research Initiative Fund. P.Z. acknowledges the support from the AI2050 Early Career Fellowship by Schmidt Sciences.
The authors thank Xunhua Zhao for valuable discussions.

\bibliography{references}

@article{gong_predictive_2025,
  title = {A Predictive Machine Learning Force-Field Framework for Liquid Electrolyte Development},
  author = {Gong, Sheng and Zhang, Yumin and Mu, Zhenliang and Pu, Zhichen and Wang, Hongyi and Han, Xu and Yu, Zhiao and Chen, Mengyi and Zheng, Tianze and Wang, Zhi and Chen, Lifei and Yang, Zhenze and Wu, Xiaojie and Shi, Shaochen and Gao, Weihao and Yan, Wen and Xiang, Liang},
  year = 2025,
  journal = {Nat. Mach. Intell.},
  volume = {7},
  number = {4},
  pages = {543--552},
  issn = {2522-5839},
  doi = {10.1038/s42256-025-01009-7}
}

@article{tissandier_protons_1998,
  title = {The Proton's Absolute Aqueous Enthalpy and {Gibbs} Free Energy of Solvation from Cluster-Ion Solvation Data},
  author = {Tissandier, Michael D. and Cowen, Kenneth A. and Feng, Wan Yong and Gundlach, Ellen and Cohen, Michael H. and Earhart, Alan D. and Coe, James V. and Tuttle, Thomas R.},
  year = {1998},
  journal = {J. Phys. Chem. A},
  volume = {102},
  number = {40},
  pages = {7787--7794},
  issn = {1089-5639, 1520-5215},
  doi = {10.1021/jp982638r}
}

@article{zhan_protons_2001,
  title = {Absolute Hydration Free Energy of the Proton from First-Principles Electronic Structure Calculations},
  author = {Zhan, Chang-Guo and Dixon, David A.},
  year = {2001},
  journal = {J. Phys. Chem. A},
  volume = {105},
  number = {51},
  pages = {11534--11540},
  issn = {1089-5639, 1520-5215},
  doi = {10.1021/jp012536s}
}

@article{fornari_pcet-ph_2021,
  title = {A Computational Protocol Combining {DFT} and Cheminformatics for Prediction of {pH}-Dependent Redox Potentials},
  author = {Fornari, Rocco Peter and {de Silva}, Piotr},
  year = {2021},
  journal = {Molecules},
  volume = {26},
  number = {13},
  pages = {3978},
  issn = {1420-3049},
  doi = {10.3390/molecules26133978}
}

@article{huynh_pcet-et_2016,
  title = {Quinone 1 e-- and 2 e--/2 {H}+ Reduction Potentials: Identification and Analysis of Deviations from Systematic Scaling Relationships},
  author = {Huynh, Mioy T. and Anson, Colin W. and Cavell, Andrew C. and Stahl, Shannon S. and {Hammes-Schiffer}, Sharon},
  year = {2016},
  journal = {J. Am. Chem. Soc.},
  volume = {138},
  number = {49},
  pages = {15903--15910},
  issn = {0002-7863},
  doi = {10.1021/jacs.6b05797}
}

@article{li_lewis-base_2022,
  title = {Redox-Tunable {Lewis} Bases for Electrochemical Carbon Dioxide Capture},
  author = {Li, Xing and Zhao, Xunhua and Liu, Yuanyue and Hatton, T. Alan and Liu, Yayuan},
  year = {2022},
  journal = {Nature Energy},
  volume = {7},
  number = {11},
  pages = {1065--1075},
  issn = {2058-7546},
  doi = {10.1038/s41560-022-01137-z}
}

@article{wedege_ph-exptl_2016,
  title = {Organic Redox Species in Aqueous Flow Batteries: Redox Potentials, Chemical Stability and Solubility},
  author = {Wedege, Kristina and Dra{\v z}evi{\'c}, Emil and Konya, Denes and Bentien, Anders},
  year = {2016},
  journal = {Sci. Rep.},
  volume = {6},
  number = {1},
  pages = {39101},
  issn = {2045-2322},
  doi = {10.1038/srep39101}
}

@article{hohenberg_hk_1964,
  title = {Inhomogeneous Electron Gas},
  author = {Hohenberg, P. and Kohn, W.},
  year = {1964},
  journal = {Phys. Rev.},
  volume = {136},
  number = {3B},
  pages = {B864-B871},
  issn = {0031-899X},
  doi = {10.1103/PhysRev.136.B864},
}

@article{kohn_ks_1965,
  title = {Self-Consistent Equations Including Exchange and Correlation Effects},
  author = {Kohn, W. and Sham, L. J.},
  year = {1965},
  journal = {Phys. Rev.},
  volume = {140},
  number = {4A},
  pages = {A1133-A1138},
  issn = {0031-899X},
  doi = {10.1103/PhysRev.140.A1133},
}

@article{mardirossian_wb97mv_2016,
  title = {{$\omega$B97M-V}: A Combinatorially Optimized, Range-Separated Hybrid, Meta-{GGA} Density Functional with {VV10} Nonlocal Correlation},
  author = {Mardirossian, Narbe and {Head-Gordon}, Martin},
  year = {2016},
  journal = {J. Chem. Phys.},
  volume = {144},
  number = {21},
  pages = {214110},
  issn = {0021-9606, 1089-7690},
  doi = {10.1063/1.4952647}
}

@article{weigend_def2_2005,
  title = {Balanced Basis Sets of Split Valence, Triple Zeta Valence and Quadruple Zeta Valence Quality for {H} to {Rn}: Design and Assessment of Accuracy},
  author = {Weigend, Florian and Ahlrichs, Reinhart},
  year = {2005},
  journal = {Phys. Chem. Chem. Phys.},
  volume = {7},
  number = {18},
  pages = {3297},
  issn = {1463-9076, 1463-9084},
  doi = {10.1039/b508541a}
}

@article{rappoport_def2D_2010,
  title = {Property-Optimized {Gaussian} Basis Sets for Molecular Response Calculations},
  author = {Rappoport, Dmitrij and Furche, Filipp},
  year = {2010},
  journal = {J. Chem. Phys.},
  volume = {133},
  number = {13},
  pages = {134105},
  issn = {0021-9606, 1089-7690},
  doi = {10.1063/1.3484283}
}

@article{weigend_def2-jkfit_2008,
  title = {{Hartree}--{Fock} Exchange Fitting Basis Sets for {H} to {Rn} {\dag}},
  author = {Weigend, Florian},
  year = {2008},
  journal = {J. Comput. Chem.},
  volume = {29},
  number = {2},
  pages = {167--175},
  issn = {0192-8651, 1096-987X},
  doi = {10.1002/jcc.20702}
}

@article{qiu_gbas_1997,
  title = {The {GB/SA} Continuum Model for Solvation. A Fast Analytical Method for the Calculation of Approximate {Born} Radii},
  author = {Qiu, Di and Shenkin, Peter S. and Hollinger, Frank P. and Still, W. Clark},
  year = {1997},
  journal = {J. Phys. Chem. A},
  volume = {101},
  number = {16},
  pages = {3005--3014},
  issn = {1089-5639, 1520-5215},
  doi = {10.1021/jp961992r}
}

@article{tomasi_pcm_2005,
  title = {Quantum Mechanical Continuum Solvation Models},
  author = {Tomasi, Jacopo and Mennucci, Benedetta and Cammi, Roberto},
  year = {2005},
  journal = {Chem. Rev.},
  volume = {105},
  number = {8},
  pages = {2999--3094},
  issn = {0009-2665, 1520-6890},
  doi = {10.1021/cr9904009}
}

@article{marenich_smd_2009,
  title = {Universal Solvation Model Based on Solute Electron Density and on a Continuum Model of the Solvent Defined by the Bulk Dielectric Constant and Atomic Surface Tensions},
  author = {Marenich, Aleksandr V. and Cramer, Christopher J. and Truhlar, Donald G.},
  year = {2009},
  journal = {J. Phys. Chem. B},
  volume = {113},
  number = {18},
  pages = {6378--6396},
  issn = {1520-6106, 1520-5207},
  doi = {10.1021/jp810292n}
}

@article{ribeiro_smd_eg_2011,
  title = {Use of Solution-Phase Vibrational Frequencies in Continuum Models for the Free Energy of Solvation},
  author = {Ribeiro, Raphael F. and Marenich, Aleksandr V. and Cramer, Christopher J. and Truhlar, Donald G.},
  year = {2011},
  journal = {J. Phys. Chem. B},
  volume = {115},
  number = {49},
  pages = {14556--14562},
  issn = {1520-6106, 1520-5207},
  doi = {10.1021/jp205508z}
}

@article{becke_b3_1993,
  title = {Density-Functional Thermochemistry. {III}. The Role of Exact Exchange},
  author = {Becke, Axel D.},
  year = {1993},
  journal = {J. Chem. Phys.},
  volume = {98},
  number = {7},
  pages = {5648--5652},
  issn = {0021-9606, 1089-7690},
  doi = {10.1063/1.464913}
}

@article{lee_lyp_1988,
  title = {Development of the {Colle-Salvetti} Correlation-Energy Formula into a Functional of the Electron Density},
  author = {Lee, Chengteh and Yang, Weitao and Parr, Robert G.},
  year = {1988},
  journal = {Phys. Rev. B},
  volume = {37},
  number = {2},
  pages = {785--789},
  issn = {0163-1829},
  doi = {10.1103/PhysRevB.37.785}
}

@article{stephens_b3lyp_1994,
  title = {\emph{Ab initio} Calculation of Vibrational Absorption and Circular Dichroism Spectra Using Density Functional Force Fields},
  author = {Stephens, P. J. and Devlin, F. J. and Chabalowski, C. F. and Frisch, M. J.},
  year = {1994},
  journal = {J. Phys. Chem.},
  volume = {98},
  number = {45},
  pages = {11623--11627},
  issn = {0022-3654, 1541-5740},
  doi = {10.1021/j100096a001}
}

@article{zhao_m062x_2008,
  title = {The {M06} Suite of Density Functionals for Main Group Thermochemistry, Thermochemical Kinetics, Noncovalent Interactions, Excited States, and Transition Elements: Two New Functionals and Systematic Testing of Four {M06}-class Functionals and 12 Other Functionals},
  author = {Zhao, Yan and Truhlar, Donald G.},
  year = {2008},
  journal = {Theor. Chem. Acc.},
  volume = {120},
  number = {1},
  pages = {215--241},
  issn = {1432-2234},
  doi = {10.1007/s00214-007-0310-x}
}

@article{chai_wb97x_2008,
  title = {Systematic Optimization of Long-Range Corrected Hybrid Density Functionals},
  author = {Chai, Jeng-Da and {Head-Gordon}, Martin},
  year = {2008},
  journal = {J. Chem. Phys.},
  volume = {128},
  number = {8},
  pages = {084106},
  issn = {0021-9606, 1089-7690},
  doi = {10.1063/1.2834918}
}

@article{najibi_wb97x/m-d3bj_2018,
  title = {The Nonlocal Kernel in {van der Waals} Density Functionals as an Additive Correction: An Extensive Analysis with Special Emphasis on the {B97M-V} and {$\omega$B97M-V Approaches}},
  author = {Najibi, Asim and Goerigk, Lars},
  year = {2018},
  journal = {J. Chem. Theory Comput.},
  volume = {14},
  number = {11},
  pages = {5725--5738},
  issn = {1549-9618, 1549-9626},
  doi = {10.1021/acs.jctc.8b00842}
}

@article{grimme_d3_2010,
  title = {A Consistent and Accurate \emph{ab initio} Parametrization of Density Functional Dispersion Correction ({DFT-D}) for the 94 Elements {H-Pu}},
  author = {Grimme, Stefan and Antony, Jens and Ehrlich, Stephan and Krieg, Helge},
  year = {2010},
  journal = {J. Chem. Phys.},
  volume = {132},
  number = {15},
  pages = {154104},
  issn = {0021-9606, 1089-7690},
  doi = {10.1063/1.3382344}
}

@article{grimme_d3bj_2011,
  title = {Effect of the Damping Function in Dispersion Corrected Density Functional Theory},
  author = {Grimme, Stefan and Ehrlich, Stephan and Goerigk, Lars},
  year = {2011},
  journal = {J. Comput. Chem.},
  volume = {32},
  number = {7},
  pages = {1456--1465},
  issn = {0192-8651, 1096-987X},
  doi = {10.1002/jcc.21759}
}

@article{bannwarth_gfn2xtb_2019,
  title = {{GFN2-xTB}---An Accurate and Broadly Parametrized Self-Consistent Tight-Binding Quantum Chemical Method with Multipole Electrostatics and Density-Dependent Dispersion Contributions},
  author = {Bannwarth, Christoph and Ehlert, Sebastian and Grimme, Stefan},
  year = {2019},
  journal = {J. Chem. Theory Comput.},
  volume = {15},
  number = {3},
  pages = {1652--1671},
  issn = {1549-9618, 1549-9626},
  doi = {10.1021/acs.jctc.8b01176}
}

@article{ditchfield_631gd_1971,
  title = {Self-Consistent Molecular-Orbital Methods. {IX}. An Extended {Gaussian}-Type Basis for Molecular-Orbital Studies of Organic Molecules},
  author = {Ditchfield, R. and Hehre, W. J. and Pople, J. A.},
  year = {1971},
  journal = {J. Chem. Phys.},
  volume = {54},
  number = {2},
  pages = {724--728},
  issn = {0021-9606, 1089-7690},
  doi = {10.1063/1.1674902}
}

@article{francl_631gd_1982,
  title = {Self-Consistent Molecular Orbital Methods. {XXIII}. A Polarization-Type Basis Set for Second-Row Elements},
  author = {Francl, Michelle M. and Pietro, William J. and Hehre, Warren J. and Binkley, J. Stephen and Gordon, Mark S. and DeFrees, Douglas J. and Pople, John A.},
  year = {1982},
  journal = {J. Chem. Phys.},
  volume = {77},
  number = {7},
  pages = {3654--3665},
  issn = {0021-9606, 1089-7690},
  doi = {10.1063/1.444267}
}

@article{gordon_631gd_1982,
  title = {Self-Consistent Molecular-Orbital Methods. 22. Small Split-Valence Basis Sets for Second-Row Elements},
  author = {Gordon, Mark S. and Binkley, J. Stephen and Pople, John A. and Pietro, William J. and Hehre, Warren J.},
  year = {1982},
  journal = {J. Am. Chem. Soc.},
  volume = {104},
  number = {10},
  pages = {2797--2803},
  issn = {0002-7863, 1520-5126},
  doi = {10.1021/ja00374a017}
}

@article{hariharan_631gd_1973,
  title = {The Influence of Polarization Functions on Molecular Orbital Hydrogenation Energies},
  author = {Hariharan, P. C. and Pople, J. A.},
  year = {1973},
  journal = {Theoret. Chim. Acta},
  volume = {28},
  number = {3},
  pages = {213--222},
  issn = {0040-5744, 1432-2234},
  doi = {10.1007/BF00533485}
}

@article{hehre_631gd_1972,
  title = {Self---Consistent Molecular Orbital Methods. {XII}. Further Extensions of {Gaussian}---Type Basis Sets for Use in Molecular Orbital Studies of Organic Molecules},
  author = {Hehre, W. J. and Ditchfield, R. and Pople, J. A.},
  year = {1972},
  journal = {J. Chem. Phys.},
  volume = {56},
  number = {5},
  pages = {2257--2261},
  issn = {0021-9606, 1089-7690},
  doi = {10.1063/1.1677527}
}

@article{cizek_cc_1966,
  title = {On the Correlation Problem in Atomic and Molecular Systems. Calculation of Wavefunction Components in {Ursell}-Type Expansion Using Quantum-Field Theoretical Methods},
  author = {{\v C}{\'i}{\v z}ek, Ji{\v r}{\'i}},
  year = {1966},
  journal = {J. Chem. Phys.},
  volume = {45},
  number = {11},
  pages = {4256--4266},
  issn = {0021-9606, 1089-7690},
  doi = {10.1063/1.1727484}
}

@article{purvis_cc_1982,
  title = {A Full Coupled-Cluster Singles and Doubles Model: The Inclusion of Disconnected Triples},
  author = {Purvis, George D. and Bartlett, Rodney J.},
  year = {1982},
  journal = {J. Chem. Phys.},
  volume = {76},
  number = {4},
  pages = {1910--1918},
  issn = {0021-9606, 1089-7690},
  doi = {10.1063/1.443164}
}

@article{raghavachari_cc_1989,
  title = {A Fifth-Order Perturbation Comparison of Electron Correlation Theories},
  author = {Raghavachari, Krishnan and Trucks, Gary W. and Pople, John A. and {Head-Gordon}, Martin},
  year = {1989},
  journal = {Chem. Phys. Lett.},
  volume = {157},
  number = {6},
  pages = {479--483},
  issn = {00092614},
  doi = {10.1016/S0009-2614(89)87395-6}
}

@article{guo_dlpno_2018,
  title = {Communication: An Improved Linear Scaling Perturbative Triples Correction for the Domain Based Local Pair-Natural Orbital Based Singles and Doubles Coupled Cluster Method [{DLPNO-CCSD(T)}]},
  author = {Guo, Yang and Riplinger, Christoph and Becker, Ute and Liakos, Dimitrios G. and Minenkov, Yury and Cavallo, Luigi and Neese, Frank},
  year = {2018},
  journal = {J. Chem. Phys.},
  volume = {148},
  number = {1},
  pages = {011101},
  issn = {0021-9606, 1089-7690},
  doi = {10.1063/1.5011798}
}

@article{riplinger_dlpno_2013,
  title = {An Efficient and near Linear Scaling Pair Natural Orbital Based Local Coupled Cluster Method},
  author = {Riplinger, Christoph and Neese, Frank},
  year = {2013},
  journal = {J. Chem. Phys.},
  volume = {138},
  number = {3},
  pages = {034106},
  issn = {0021-9606, 1089-7690},
  doi = {10.1063/1.4773581}
}

@article{riplinger_dlpno_2016,
  title = {Sparse Maps---A Systematic Infrastructure for Reduced-Scaling Electronic Structure Methods. {II}. Linear Scaling Domain Based Pair Natural Orbital Coupled Cluster Theory},
  author = {Riplinger, Christoph and Pinski, Peter and Becker, Ute and Valeev, Edward F. and Neese, Frank},
  year = {2016},
  journal = {J. Chem. Phys.},
  volume = {144},
  number = {2},
  pages = {024109},
  issn = {0021-9606, 1089-7690},
  doi = {10.1063/1.4939030}
}

@article{saitow_dlpno_2017,
  title = {A New Near-Linear Scaling, Efficient and Accurate, Open-Shell Domain-Based Local Pair Natural Orbital Coupled Cluster Singles and Doubles Theory},
  author = {Saitow, Masaaki and Becker, Ute and Riplinger, Christoph and Valeev, Edward F. and Neese, Frank},
  year = {2017},
  journal = {J. Chem. Phys.},
  volume = {146},
  number = {16},
  pages = {164105},
  issn = {0021-9606, 1089-7690},
  doi = {10.1063/1.4981521}
}

@article{adler_f12_2007,
  title = {A Simple and Efficient {CCSD(T)-F12} Approximation},
  author = {Adler, Thomas B. and Knizia, Gerald and Werner, Hans-Joachim},
  year = {2007},
  journal = {J. Chem. Phys.},
  volume = {127},
  number = {22},
  pages = {221106},
  issn = {0021-9606, 1089-7690},
  doi = {10.1063/1.2817618}
}

@misc{levine_omol25_2025,
  title = {The Open Molecules 2025 ({OMol25}) Dataset, Evaluations, and Models},
  author = {Levine, Daniel S. and others},
  year = {2025},
  number = {arXiv:2505.08762},
  eprint = {2505.08762},
  primaryclass = {physics},
  doi = {10.48550/arXiv.2505.08762},
  archiveprefix = {arXiv}
}

@article{sun_pyscf_2020,
  title = {Recent Developments in the {P\textsc{y}SCF} Program Package},
  author = {Sun, Qiming and others},
  year = {2020},
  journal = {J. Chem. Phys.},
  volume = {153},
  number = {2},
  pages = {024109},
  issn = {0021-9606, 1089-7690},
  doi = {10.1063/5.0006074}
}

@article{li_gpu4pyscf_2025,
  title = {Introducing {GPU} Acceleration into the {Python}-Based Simulations of Chemistry Framework},
  author = {Li, Rui and Sun, Qiming and Zhang, Xing and Chan, Garnet Kin-Lic},
  year = {2025},
  journal = {J. Phys. Chem. A},
  volume = {129},
  number = {5},
  pages = {1459--1468},
  issn = {1089-5639},
  doi = {10.1021/acs.jpca.4c05876}
}

@article{wu_gpu4pyscf_2025,
  title = {Enhancing {GPU}-Acceleration in the {Python}-Based Simulations of Chemistry Frameworks},
  author = {Wu, Xiaojie and Sun, Qiming and Pu, Zhichen and Zheng, Tianze and Ma, Wenzhi and Yan, Wen and Xia, Yu and Wu, Zhengxiao and Huo, Mian and Li, Xiang and Ren, Weiluo and Gong, Sheng and Zhang, Yumin and Gao, Weihao},
  year = {2025},
  journal = {WIREs Comput. Mol. Sci.},
  volume = {15},
  number = {2},
  pages = {e70008},
  issn = {1759-0876, 1759-0884},
  doi = {10.1002/wcms.70008}
}

@article{pracht_crest_2024,
  title = {{CREST}---A Program for the Exploration of Low-Energy Molecular Chemical Space},
  author = {Pracht, Philipp and Grimme, Stefan and Bannwarth, Christoph and Bohle, Fabian and Ehlert, Sebastian and Feldmann, Gereon and Gorges, Johannes and M{\"u}ller, Marcel and Neudecker, Tim and Plett, Christoph and Spicher, Sebastian and Steinbach, Pit and Weso{\l}owski, Patryk A. and Zeller, Felix},
  year = {2024},
  journal = {J. Chem. Phys.},
  volume = {160},
  number = {11},
  pages = {114110},
  issn = {0021-9606, 1089-7690},
  doi = {10.1063/5.0197592}
}

@article{hermes_sella_2021,
  title = {Geometry Optimization Speedup through a Geodesic Approach to Internal Coordinates},
  author = {Hermes, Eric D. and Sargsyan, Khachik and Najm, Habib N. and Z{\'a}dor, Judit},
  year = {2021},
  journal = {J. Chem. Phys.},
  volume = {155},
  number = {9},
  pages = {094105},
  issn = {0021-9606, 1089-7690},
  doi = {10.1063/5.0060146}
}

@misc{landrum_rdkit_2025,
    author = {Landrum, Gregory A.},
    title = {{RDKit}: Open-Source Cheminformatics},
    url = {https://www.rdkit.org/},
    doi = {10.5281/zenodo.16996017},
    year = {2025},
}

@article{chen_universal_2022,
  title = {A Universal Graph Deep Learning Interatomic Potential for the Periodic Table},
  author = {Chen, Chi and Ong, Shyue Ping},
  year = 2022,
  journal = {Nat. Comput. Sci.},
  volume = {2},
  number = {11},
  pages = {718--728},
  issn = {2662-8457},
  doi = {10.1038/s43588-022-00349-3}
}

@article{deng_chgnet_2023,
  title = {{CHGNet} as a Pretrained Universal Neural Network Potential for Charge-Informed Atomistic Modelling},
  author = {Deng, Bowen and Zhong, Peichen and Jun, KyuJung and Riebesell, Janosh and Han, Kevin and Bartel, Christopher J. and Ceder, Gerbrand},
  year = 2023,
  journal = {Nat. Mach. Intell.},
  volume = {5},
  number = {9},
  pages = {1031--1041},
  issn = {2522-5839},
  doi = {10.1038/s42256-023-00716-3}
}

@article{zhou_carbon_2024,
    title = {Carbon dioxide capture from open air using covalent organic frameworks},
    volume = {635},
    issn = {0028-0836, 1476-4687},
    url = {https://www.nature.com/articles/s41586-024-08080-x},
    doi = {10.1038/s41586-024-08080-x},
    number = {8037},
    urldate = {2025-10-06},
    journal = {Nature},
    author = {Zhou, Zihui and Ma, Tianqiong and Zhang, Heyang and Chheda, Saumil and Li, Haozhe and Wang, Kaiyu and Ehrling, Sebastian and Giovine, Raynald and Li, Chuanshuai and Alawadhi, Ali H. and Abduljawad, Marwan M. and Alawad, Majed O. and Gagliardi, Laura and Sauer, Joachim and Yaghi, Omar M.},
    year = {2024},
    pages = {96--101},
}

@article{chu_carbon_2009,
    title = {Carbon Capture and Sequestration},
    volume = {325},
    issn = {0036-8075, 1095-9203},
    url = {https://www.science.org/doi/10.1126/science.1181637},
    doi = {10.1126/science.1181637},
    language = {english},
    number = {5948},
    urldate = {2025-10-06},
    journal = {Science},
    author = {Chu, Steven},
    month = sep,
    year = {2009},
    pages = {1599--1599},
}

@article{lin_scalable_2021,
  title = {A Scalable Metal-Organic Framework as a Durable Physisorbent for Carbon Dioxide Capture},
  author = {Lin, Jian-Bin and Nguyen, Tai T. T. and Vaidhyanathan, Ramanathan and Burner, Jake and Taylor, Jared M. and Durekova, Hana and Akhtar, Farid and Mah, Roger K. and {Ghaffari-Nik}, Omid and Marx, Stefan and Fylstra, Nicholas and Iremonger, Simon S. and Dawson, Karl W. and Sarkar, Partha and Hovington, Pierre and Rajendran, Arvind and Woo, Tom K. and Shimizu, George K. H.},
  year = 2021,
  journal = {Science},
  volume = {374},
  number = {6574},
  pages = {1464--1469},
  issn = {0036-8075, 1095-9203},
  doi = {10.1126/science.abi7281}
}

@article{siegelman_porous_2021,
  title = {Porous Materials for Carbon Dioxide Separations},
  author = {Siegelman, Rebecca L. and Kim, Eugene J. and Long, Jeffrey R.},
  year = 2021,
  journal = {Nat. Mater.},
  volume = {20},
  number = {8},
  pages = {1060--1072},
  issn = {1476-1122, 1476-4660},
  doi = {10.1038/s41563-021-01054-8}
}

@article{diederichsen_electrochemical_2022,
  title = {Electrochemical Methods for Carbon Dioxide Separations},
  author = {Diederichsen, Kyle M. and Sharifian, Rezvan and Kang, Jin Soo and Liu, Yayuan and Kim, Seoni and Gallant, Betar M. and Vermaas, David and Hatton, T. Alan},
  year = 2022,
  journal = {Nat. Rev. Methods Primers},
  volume = {2},
  number = {1},
  pages = {68},
  issn = {2662-8449},
  doi = {10.1038/s43586-022-00148-0}
}

@article{sharifian_electrochemical_2021,
  title = {Electrochemical Carbon Dioxide Capture to Close the Carbon Cycle},
  author = {Sharifian, R. and Wagterveld, R. M. and Digdaya, I. A. and Xiang, C. and Vermaas, D. A.},
  year = 2021,
  journal = {Energy Environ. Sci.},
  volume = {14},
  number = {2},
  pages = {781--814},
  issn = {1754-5692, 1754-5706},
  doi = {10.1039/D0EE03382K}
}

@article{voskian_faradaic_2019,
  title = {Faradaic Electro-Swing Reactive Adsorption for {{CO}}{\textsubscript{2}} Capture},
  author = {Voskian, Sahag and Hatton, T. Alan},
  year = 2019,
  journal = {Energy Environ. Sci.},
  volume = {12},
  number = {12},
  pages = {3530--3547},
  issn = {1754-5692, 1754-5706},
  doi = {10.1039/C9EE02412C}
}

@article{diederichsen_solventfree_2022,
  title = {Toward Solvent-Free Continuous-Flow Electrochemically Mediated Carbon Capture with High-Concentration Liquid Quinone Chemistry},
  author = {Diederichsen, Kyle M. and Liu, Yayuan and Ozbek, Nil and Seo, Hyowon and Hatton, T. Alan},
  year = 2022,
  journal = {Joule},
  volume = {6},
  number = {1},
  pages = {221--239},
  issn = {25424351},
  doi = {10.1016/j.joule.2021.12.001}
}

@article{li_accurate_2025,
  title = {Accurate {QM/MM} Molecular Dynamics for Periodic Systems in {GPU4P\textsc{y}SCF} with Applications to Enzyme Catalysis},
  author = {Li, Chenghan and Chan, Garnet Kin-Lic},
  year = 2025,
  journal = {J. Chem. Theory Comput.},
  volume = {21},
  number = {2},
  pages = {803--816},
  issn = {1549-9618, 1549-9626},
  doi = {10.1021/acs.jctc.4c01698}
}

@misc{yang_mattersim_2024,
    title = {{MatterSim}: A Deep Learning Atomistic Model Across Elements, Temperatures and Pressures},
    shorttitle = {{MatterSim}},
    url = {http://arxiv.org/abs/2405.04967},
    doi = {10.48550/arXiv.2405.04967},
    urldate = {2025-10-06},
    publisher = {arXiv},
    author = {Yang, Han and Hu, Chenxi and Zhou, Yichi and Liu, Xixian and Shi, Yu and Li, Jielan and Li, Guanzhi and Chen, Zekun and Chen, Shuizhou and Zeni, Claudio and Horton, Matthew and Pinsler, Robert and Fowler, Andrew and Zügner, Daniel and Xie, Tian and Smith, Jake and Sun, Lixin and Wang, Qian and Kong, Lingyu and Liu, Chang and Hao, Hongxia and Lu, Ziheng},
    month = may,
    year = {2024},
    note = {arXiv:2405.04967 [cond-mat]},
}

@misc{rhodes_orb-v3_2025,
    title = {{Orb-v3}: atomistic simulation at scale},
    shorttitle = {Orb-v3},
    url = {http://arxiv.org/abs/2504.06231},
    doi = {10.48550/arXiv.2504.06231},
    urldate = {2025-10-06},
    publisher = {arXiv},
    author = {Rhodes, Benjamin and Vandenhaute, Sander and Šimkus, Vaidotas and Gin, James and Godwin, Jonathan and Duignan, Tim and Neumann, Mark},
    month = apr,
    year = {2025},
    note = {arXiv:2504.06231 [cond-mat]},
}

@misc{fu_learning_2025,
    title = {Learning Smooth and Expressive Interatomic Potentials for Physical Property Prediction},
    url = {http://arxiv.org/abs/2502.12147},
    doi = {10.48550/arXiv.2502.12147},
    urldate = {2025-10-06},
    publisher = {arXiv},
    author = {Fu, Xiang and Wood, Brandon M. and Barroso-Luque, Luis and Levine, Daniel S. and Gao, Meng and Dzamba, Misko and Zitnick, C. Lawrence},
    month = apr,
    year = {2025},
    note = {arXiv:2502.12147 [physics]},
}

@misc{zhang_graph_2025,
    title = {A Graph Neural Network for the Era of Large Atomistic Models},
    url = {http://arxiv.org/abs/2506.01686},
    doi = {10.48550/arXiv.2506.01686},
    urldate = {2025-10-06},
    publisher = {arXiv},
    author = {Zhang, Duo and Peng, Anyang and Cai, Chun and Li, Wentao and Zhou, Yuanchang and Zeng, Jinzhe and Guo, Mingyu and Zhang, Chengqian and Li, Bowen and Jiang, Hong and Zhu, Tong and Jia, Weile and Zhang, Linfeng and Wang, Han},
    month = jun,
    year = {2025},
    note = {arXiv:2506.01686 [physics]},
}

@article{kim_dataefficient_2025,
  title = {Data-Efficient Multifidelity Training for High-Fidelity Machine Learning Interatomic Potentials},
  author = {Kim, Jaesun and Kim, Jisu and Kim, Jaehoon and Lee, Jiho and Park, Yutack and Kang, Youngho and Han, Seungwu},
  year = 2025,
  journal = {J. Am. Chem. Soc.},
  volume = {147},
  number = {1},
  pages = {1042--1054},
  issn = {0002-7863, 1520-5126},
  doi = {10.1021/jacs.4c14455}
}

@article{baik_computing_2002,
  title = {Computing Redox Potentials in Solution: Density Functional Theory as A Tool for Rational Design of Redox Agents},
  author = {Baik, Mu-Hyun and Friesner, Richard A.},
  year = 2002,
  journal = {J. Phys. Chem. A},
  volume = {106},
  number = {32},
  pages = {7407--7412},
  issn = {1089-5639, 1520-5215},
  doi = {10.1021/jp025853n}
}

@article{horton_accelerated_2025,
  title = {Accelerated Data-Driven Materials Science with the {Materials Project}},
  author = {Horton, Matthew K. and others},
  year = 2025,
  journal = {Nat. Mater.},
  volume = {24},
  number = {10},
  pages = {1522--1532},
  issn = {1476-1122, 1476-4660},
  doi = {10.1038/s41563-025-02272-0}
}

@article{bochkarev_graph_2024,
  title = {Graph Atomic Cluster Expansion for Semilocal Interactions beyond Equivariant Message Passing},
  author = {Bochkarev, Anton and Lysogorskiy, Yury and Drautz, Ralf},
  year = 2024,
  journal = {Phys. Rev. X},
  volume = {14},
  number = {2},
  pages = {021036},
  issn = {2160-3308},
  doi = {10.1103/PhysRevX.14.021036}
}

@article{bremond_rangeseparated_2018,
  title = {Range-Separated Double-Hybrid Functional from Nonempirical Constraints},
  author = {Br{\'e}mond, {\'E}ric and Savarese, Marika and {P{\'e}rez-Jim{\'e}nez}, {\'A}ngel Jos{\'e} and {Sancho-Garc{\'i}a}, Juan Carlos and Adamo, Carlo},
  year = {2018},
  journal = {J. Chem. Theory Comput.},
  volume = {14},
  number = {8},
  pages = {4052--4062},
  issn = {1549-9618, 1549-9626},
  doi = {10.1021/acs.jctc.8b00261}
}

@article{wang_materials_2025,
  title = {Materials Design and Assessment of Redox-Mediated Flow Cell Systems for Enhanced Energy Storage and Conversion},
  author = {Wang, Zhiyu and Jing, Yan and Wang, Qing},
  year = 2025,
  journal = {Adv. Mater.},
  pages = {e09991},
  issn = {0935-9648, 1521-4095},
  doi = {10.1002/adma.202509991}
}

@article{sola_menshutkin_1991,
  title = {Analysis of Solvent Effects on the {Menshutkin} Reaction},
  author = {Sola, Miquel and Lledos, Agusti and Duran, Miquel and Bertran, Juan and Abboud, Jose Luis M.},
  year = {1991},
  journal = {J. Am. Chem. Soc.},
  volume = {113},
  number = {8},
  pages = {2873--2879},
  issn = {0002-7863, 1520-5126},
  doi = {10.1021/ja00008a013}
}

@misc{kalai_why_2025,
    title = {Why Language Models Hallucinate},
    url = {http://arxiv.org/abs/2509.04664},
    doi = {10.48550/arXiv.2509.04664},
    language = {english},
    urldate = {2025-10-12},
    publisher = {arXiv},
    author = {Kalai, Adam Tauman and Nachum, Ofir and Vempala, Santosh S. and Zhang, Edwin},
    month = sep,
    year = {2025},
    note = {arXiv:2509.04664 [cs]},
}

@article{trasatti_she_1986,
  title = {The Absolute Electrode Potential: An Explanatory Note (Recommendations 1986)},
  author = {Trasatti, Sergio},
  year = {1986},
  journal = {J. Electroanal. Chem. Interfacial Electrochem.},
  volume = {209},
  number = {2},
  pages = {417--428}
}

@article{morris_born-fajans-haber_1969,
    title = {The {Born-Fajans-Haber} Correlation},
    volume = {224},
    copyright = {http://www.springer.com/tdm},
    issn = {0028-0836, 1476-4687},
    url = {https://www.nature.com/articles/224950a0},
    doi = {10.1038/224950a0},
    language = {english},
    number = {5223},
    urldate = {2025-10-13},
    journal = {Nature},
    author = {Morris, D. F. C. and Short, E. L.},
    month = dec,
    year = {1969},
    pages = {950--952},
}

@misc{batatia_mace_2023,
    title = {{MACE}: Higher Order Equivariant Message Passing Neural Networks for Fast and Accurate Force Fields},
    shorttitle = {{MACE}},
    url = {http://arxiv.org/abs/2206.07697},
    doi = {10.48550/arXiv.2206.07697},
    urldate = {2025-09-01},
    publisher = {arXiv},
    author = {Batatia, Ilyes and Kovács, Dávid Péter and Simm, Gregor N. C. and Ortner, Christoph and Csányi, Gábor},
    month = jan,
    year = {2023},
    note = {arXiv:2206.07697 [stat]},
}

@article{jin_ph_2020,
  title = {{pH} Swing Cycle for {CO\textsubscript{2}} Capture Electrochemically Driven through Proton-Coupled Electron Transfer},
  author = {Jin, Shijian and Wu, Min and Gordon, Roy G. and Aziz, Michael J. and Kwabi, David G.},
  year = 2020,
  journal = {Energy Environ. Sci.},
  volume = {13},
  number = {10},
  pages = {3706--3722},
  issn = {1754-5692, 1754-5706},
  doi = {10.1039/D0EE01834A}
}

@article{xie_lowenergy_2020,
  title = {Low-Energy Electrochemical Carbon Dioxide Capture Based on a Biological Redox Proton Carrier},
  author = {Xie, Heping and Jiang, Wenchuan and Liu, Tao and Wu, Yifan and Wang, Yufei and Chen, Bin and Niu, Dawen and Liang, Bin},
  year = 2020,
  journal = {Cell Rep. Phys. Sci.},
  volume = {1},
  number = {5},
  pages = {100046},
  issn = {26663864},
  doi = {10.1016/j.xcrp.2020.100046}
}

@article{jing_electrochemically_2024,
  title = {Electrochemically Induced {CO\textsubscript{2}} Capture Enabled by Aqueous Quinone Flow Chemistry},
  author = {Jing, Yan and Amini, Kiana and Xi, Dawei and Jin, Shijian and Alfaraidi, Abdulrahman M. and Kerr, Emily F. and Gordon, Roy G. and Aziz, Michael J.},
  year = 2024,
  journal = {ACS Energy Lett.},
  volume = {9},
  number = {7},
  pages = {3526--3535},
  issn = {2380-8195, 2380-8195},
  doi = {10.1021/acsenergylett.4c01235}
}

@article{huskinson_metal-free_2014,
    title = {A metal-free organic–inorganic aqueous flow battery},
    volume = {505},
    copyright = {http://www.springer.com/tdm},
    issn = {0028-0836, 1476-4687},
    url = {https://www.nature.com/articles/nature12909},
    doi = {10.1038/nature12909},
    language = {english},
    number = {7482},
    urldate = {2025-10-15},
    journal = {Nature},
    author = {Huskinson, Brian and Marshak, Michael P. and Suh, Changwon and Er, Süleyman and Gerhardt, Michael R. and Galvin, Cooper J. and Chen, Xudong and Aspuru-Guzik, Alán and Gordon, Roy G. and Aziz, Michael J.},
    month = jan,
    year = {2014},
    pages = {195--198},
}

@article{liang_universal_2017,
  title = {Universal Quinone Electrodes for Long Cycle Life Aqueous Rechargeable Batteries},
  author = {Liang, Yanliang and Jing, Yan and Gheytani, Saman and Lee, Kuan-Yi and Liu, Ping and Facchetti, Antonio and Yao, Yan},
  year = 2017,
  journal = {Nat. Mater.},
  volume = {16},
  number = {8},
  pages = {841--848},
  issn = {1476-1122, 1476-4660},
  doi = {10.1038/nmat4919}
}

@article{zhang_decoupled_2021,
  title = {Decoupled Redox Catalytic Hydrogen Production with a Robust Electrolyte-Borne Electron and Proton Carrier},
  author = {Zhang, Feifei and Zhang, Hang and Salla, Manohar and Qin, Ning and Gao, Mengqi and Ji, Ya and Huang, Shiqiang and Wu, Sisi and Zhang, Ruifeng and Lu, Zhouguang and Wang, Qing},
  year = 2021,
  journal = {J. Am. Chem. Soc.},
  volume = {143},
  number = {1},
  pages = {223--231},
  issn = {0002-7863, 1520-5126},
  doi = {10.1021/jacs.0c09510}
}

@article{gagne_ferrocene_1980,
  title = {Ferrocene as an Internal Standard for Electrochemical Measurements},
  author = {Gagne, Robert R. and Koval, Carl A. and Lisensky, George C.},
  year = 1980,
  journal = {Inorg. Chem.},
  volume = {19},
  number = {9},
  pages = {2854--2855},
  issn = {0020-1669, 1520-510X},
  doi = {10.1021/ic50211a080}
}

@misc{yuan_foundation_2025,
    title = {Foundation {Models} for {Atomistic} {Simulation} of {Chemistry} and {Materials}},
    url = {http://arxiv.org/abs/2503.10538},
    doi = {10.48550/arXiv.2503.10538},
    urldate = {2025-10-27},
    publisher = {arXiv},
    author = {Yuan, Eric C.-Y. and Liu, Yunsheng and Chen, Junmin and Zhong, Peichen and Raja, Sanjeev and Kreiman, Tobias and Vargas, Santiago and Xu, Wenbin and Head-Gordon, Martin and Yang, Chao and Blau, Samuel M. and Cheng, Bingqing and Krishnapriyan, Aditi and Head-Gordon, Teresa},
    month = jun,
    year = {2025},
    note = {arXiv:2503.10538 [physics]},
}

@misc{wood_uma_2025,
      title={{UMA}: A Family of Universal Models for Atoms}, 
      author={Brandon M. Wood and Misko Dzamba and Xiang Fu and Meng Gao and Muhammed Shuaibi and Luis Barroso-Luque and Kareem Abdelmaqsoud and Vahe Gharakhanyan and John R. Kitchin and Daniel S. Levine and Kyle Michel and Anuroop Sriram and Taco Cohen and Abhishek Das and Ammar Rizvi and Sushree Jagriti Sahoo and Zachary W. Ulissi and C. Lawrence Zitnick},
      year={2025},
      eprint={2506.23971},
      archivePrefix={arXiv},
      primaryClass={cs.LG},
      url={https://arxiv.org/abs/2506.23971}, 
}

@misc{vanzanten_benchmarking_2025,
    title = {Benchmarking {OMol25}-{Trained} {Models} on {Experimental} {Reduction}-{Potential} and {Electron}-{Affinity} {Data}},
    copyright = {https://creativecommons.org/licenses/by/4.0/},
    url = {https://chemrxiv.org/engage/chemrxiv/article-details/68b75742728bf9025ede9923},
    doi = {10.26434/chemrxiv-2025-3stpx},
    urldate = {2025-12-17},
    publisher = {Chemistry},
    author = {VanZanten, Sawyer and Wagen, Corin},
    month = sep,
    year = {2025},
    note = {chemrxiv-2025-3stpx}
}
\end{document}